\documentclass[reprint,prl,aps, nofootinbib]{revtex4-1}
\usepackage{natbib}
\usepackage{graphicx}
\usepackage{color}
\usepackage{dcolumn}
\usepackage{bm}
\usepackage{amssymb}
\usepackage{color,amsmath}
\usepackage{soul,xcolor} 
\setstcolor{red} 
\usepackage{epstopdf}
\usepackage{longtable}
\usepackage{hyperref}
\hypersetup{
 pdfnewwindow=true, colorlinks=true,
 linkcolor=blue, anchorcolor=blue,
 citecolor=blue, filecolor=blue,
 menucolor=blue, urlcolor=blue}

\begin{document}
\title{Manipulating multi-vortex states in superconducting structures}

\author{Hryhoriy Polshyn$^{1,2,\dagger}$}
\author{Tyler Naibert$^{1}$}
\author{Raffi Budakian$^{1,3,4,5}$}

\affiliation{$^{1}$Department of Physics, University of Illinois at Urbana-Champaign, Urbana, Illinois 61801, USA}
\affiliation{$^{2}$Department of Physics, University of California, Santa Barbara CA 93106 USA}
\affiliation{$^{3}$Department of Physics, University of Waterloo, Waterloo, ON, Canada, N2L3G1}
\affiliation{$^{4}$Institute for Quantum Computing, University of Waterloo, Waterloo, ON, Canada, N2L3G1}
\affiliation{$^{5}$Perimeter Institute for Theoretical Physics, Waterloo, ON, Canada, N2L2Y5}
\affiliation{$^{\dagger}$ polshyn@ucsb.edu}

\begin{abstract}
We demonstrate a method for manipulating small ensembles of vortices in multiply-connected superconducting structures. A micron-size magnetic particle attached to the tip of a silicon cantilever is used to locally apply magnetic flux through the superconducting structure. By scanning the tip over the surface of the device, and by utilizing the dynamical coupling between the vortices and the cantilever, a high-resolution spatial map of the different vortex configurations is obtained. Moving the tip to a particular location in the map stabilizes a distinct multi-vortex configuration. Thus, the scanning of the tip over a particular trajectory in space permits non-trivial operations to be performed, such as braiding of individual vortices within a larger vortex ensemble -- a key capability required by many proposals for topological quantum computing.
\end{abstract}

\maketitle

\begin{figure*}[t]
\includegraphics[]{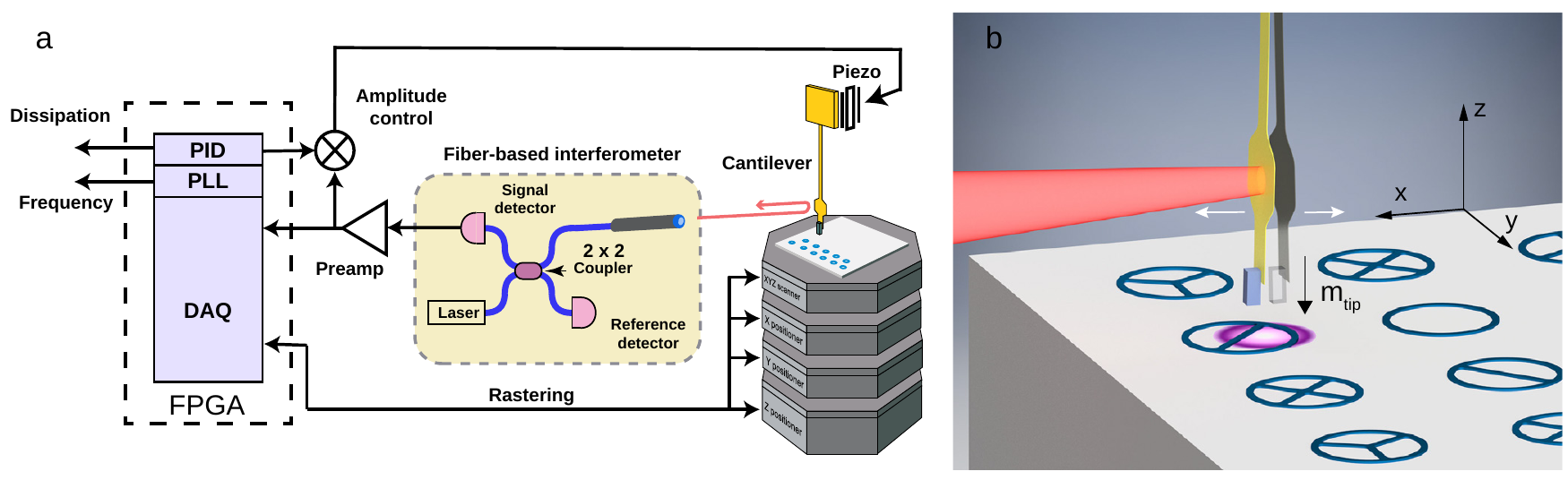}
\caption [Fig 1]{$\Phi_0$-MFM setup.(a) The schematic diagram shows the main components of the setup, including the sample mounted on the stack of nano-positioners, the silicon cantilever with magnetic particle, the fiber-based laser interfered for signal detection, and the FPGA-based electronics for signal processing. (b) Close-up view of the cantilever above superconducting structures. The inhomogenious magnetic field  produced by the magnetic particle is shown as a purple region.}
\label{Fig1}
\end{figure*} 

\begin{figure*}[t] 
\includegraphics[]{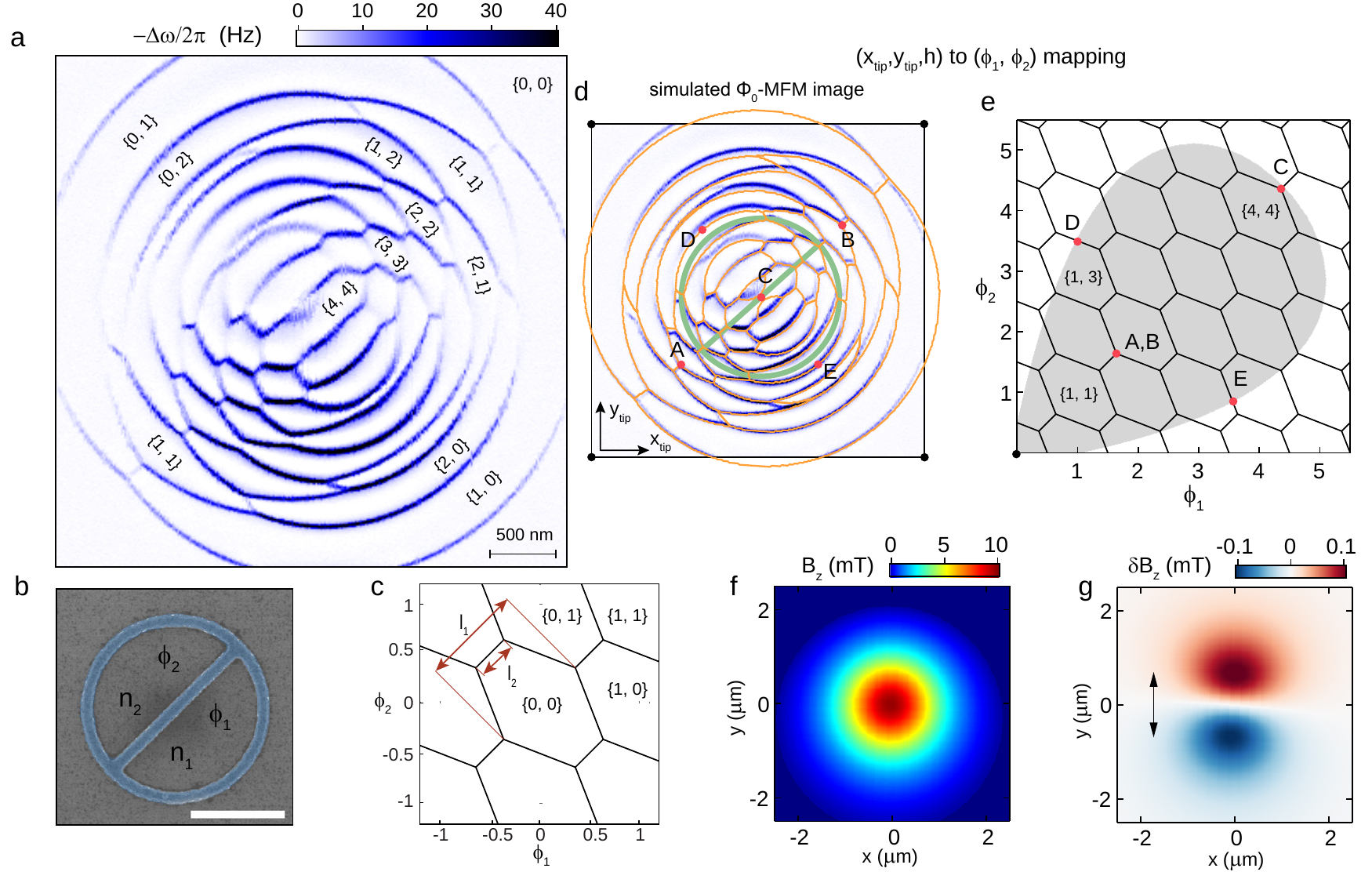}
\caption[] {Vortex states in a ring with a crossbar. (a) $\Phi_0$-MFM image of Ring~1  obtained for $h=800$~nm at 0.915~K, close to the superconducting  transition of the sample $T_c\approx 0.95$~K. (b) Scanning electron microscopy (SEM) image of Ring~1.  (c)  Stability diagram of the vortex states supported by the ring with a crossbar geometry. The number of vortices in each sector is denoted by the quantum numbers $(n_1,n_2)$ and the flux values by $(\phi_1,\phi_2)$, as shown in (b). The coupling parameter $\beta$ may be related to the ratio of the major and minor vertex separations $l_1$ and $l_2$ of the hexagonal stability region corresponding to a particular vortex state $\beta=l_2/l_1$. 
(d) and (e) show the mapping of the vortex states from ($x_{tip}, y_{tip},h)$ plane of the $\Phi_0$-MFM scan (d)  to the ($\phi_1$, $\phi_2$) plane (e). The shaded region indicated in (e) corresponds to the range of flux values that are accessed for the scan shown in (d).  Points A-E indicated in (d) and (e) are presented to illustrate the mapping between different points in position and vortex configuration space. (f) Distribution of the magnetic field $B_z(x,y)$ 800~nm below the tip. (g)Magnetic field modulation $\delta B_z (x,y)$, generated by the tip oscillations with 5~nm peak-to-peak amplitude. The double arrow shows the oscillation direction of the cantilever. }
\label{Fig2}
\end{figure*}

\begin{figure*}[t]
\centering
\includegraphics[]{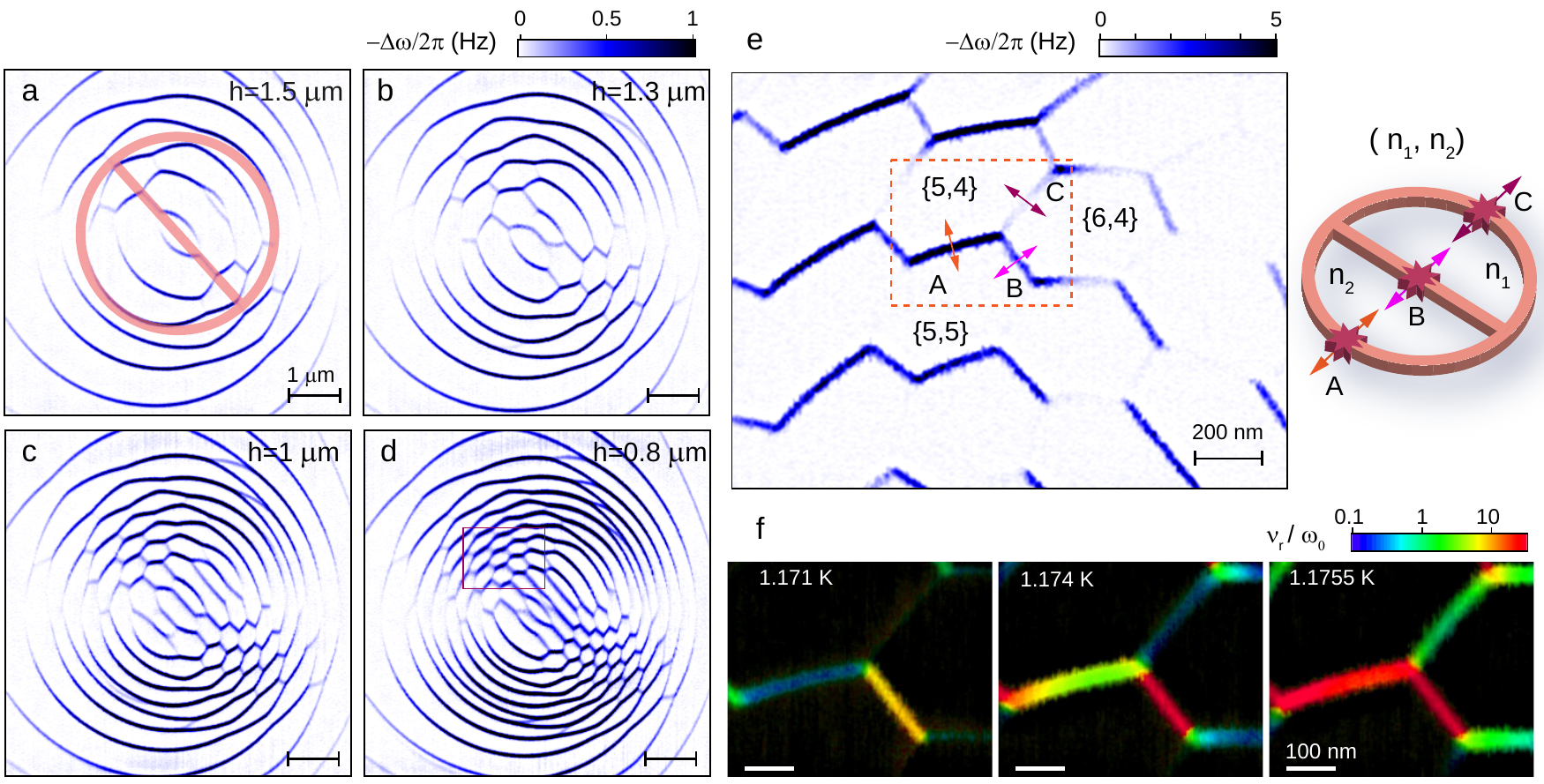}
\caption[] {(a)-(d)$\Phi_0$-MFM images of Ring~2 at several tip-surface separations. The outline of Ring~2 is shown in (a).
(e) Detailed $\Phi_0$-MFM image of the honeycomb pattern of vortex transitions measured in Ring~2 (marked by a rectangular region shown in (d)). Three selected vortex states are labeled with pairs of winding numbers $(n_1,n_2)$. The transition rates  for three vortex transitions between the states $(5,4)$, $(5,5)$ and $(6,4)$ are measured at 1.171~K, 1.174~K and 1.1755~K as shown in (f). The schematic diagram next to panel (e) illustrates the three different types of vortex entry observed for this device.
}
\label{Fig3} 
\end{figure*}

The dynamics of superconducting vortices  plays an important role in many phenomena.  In bulk superconductors the motion of macroscopic number of vortices is responsible for flux flow, pinning and creep~\cite{Tinkham1996}. In mesoscopic superconducting structures, the energetics and dynamics of small numbers of vortices  gives rise to a plethora of unexpected vortex states and  mesoscopic effects~\cite{Bezryadin_nucleation_1996,Geim1997,Geim1998, Morelle2004, Grigorieva2006, Timmermans_direct_2016, embon2017imaging}, the understanding of  which often  requires numerical simulations~\cite{Peeters2000, Baelus2000, Baelus_2001, Baelus_2002}. 
Being topological defects of superconducting order parameter, vortices play a key role in such profound effects as Aharonov-Casher effect~\cite{Aharonov1984} and  Berezinskii-Kosterlitz-Thouless transition in two-dimensional superconductors~\cite{Kosterlitz_1973}. In recent years, superconducting vortices have also been proposed as resource for the implementation of topological quantum computation. It has been suggested that vortices in superconductor-topological-insulator-superconductor junctions  may host Majorana bound states (MBS) \cite{Fu2008, beenaker_2013, Ren_2019} permitting topological quantum computation.
More recently, Abrikosov vortices in iron-based superconductors have emerged as a promising platform for manipulating MBSs.
The normal state of these materials possesses the necessary topological properties, hence the cores of Abrikosov vortices could support MBSs; recent experiments provide evidence to support the existence of MBSs in iron-based superconducting vortices~\cite{zhang_multiple_2019, wang_evidence_2018, Liu_robust_2018, zhang_observation_2018}.

Considerable efforts have been devoted to developing new ways to manipulate individual vortices.
The control of vortices has been demonstrated by means of electrical currents~\cite{Kalisky2009, Embon2015, kalcheim_dynamic_2017, ji_vortex_2016, Mills2015, Silva2006},  focused laser beams~\cite{Veshchunov2016},  local  mechanical stress~\cite{Kremen2016},  local magnetic fields~\cite{Gardner2002,Straver2008, Auslaender2009}, local heating~\cite{ge_controlled_2017, ge2016nanoscale}, and also was investigated numerically~\cite{xiaoyu_manipulation_2018, Milosevic2010}.
In most of these studies, vortices were controlled one at a time, which is an excellent approach for investigating the physics of vortex pinning at the nanoscale.
On the other hand, for applications such as vortex braiding, it is essential to be able to control individual vortex \textit{states}, and to be able to evolve these states along desired trajectories within the space of possible vortex configurations.

In this letter we use a recently developed variant of magnetic force microscopy (MFM), which we refer to as $\Phi_0$-MFM~\cite{Polshyn_2017, polshyn_magnetic_2017}, to probe and manipulate individual multi-vortex states in multiply-connected mesoscopic superconducting structures. 
The devices, shown schematically in Fig. 1b, are patterned from a thin aluminum film into rings containing one, two, three and four equal-area sectors. Below the superconducting temperature of aluminum ($T_c\approx$1.2~K) and in the presence of an applied magnetic flux, these structures can host discrete multi-vortex states, described by the number of vortices in each sector.
We use the spatially inhomogeneous magnetic field produced by a micron-sized magnetic particle attached to a cantilever to access complex multi-vortex states, many of which simply can not be stabilized by a homogeneous external magnetic field. We show that as the particle is scanned over a particular structure, the transitions between different vortex sates are marked by a shift in the resonance frequency and dissipation of the cantilever, arising from the cantilever-driven transitions between states with different vortex configurations.
We show that the spatial patterns of the frequency shift  vs. position that emerge can be mapped directly to transitions between different multi-vortex states. Furthermore, we demonstrate that this mapping enables deterministic control of the multi-vortex states of the superconducting device.

The schematic diagram of $\Phi_0$-MFM microscopy setup is shown in Fig.~\ref{Fig1}a. The $\Phi_0$-MFM images were obtained using an ultra-soft silicon cantilever with a micron-size $\mathrm{SmCo_5}$ particle attached to the tip.
Both the cantilever and the magnetic moment of the $\mathrm{SmCo_5}$ particle are perpendicular to the plane of the carrier chip, as shown in Fig.~\ref{Fig1}b. 
The magnetic particle  generates a static magnetic field $B_z(x,y)$ and modulation $\delta B_z (x,y) \sin(\omega_0 t)$ (caused by cantilever oscillations at its resonant frequency $\omega_0$) in the pane of the sample.
Two cantilevers used for this work have resonance frequencies $\omega_0/2\pi\simeq$  4146~Hz and 7675~Hz and spring constants $k=$0.11~mN/m and 0.18~mN/m, respectively (for more information and fabrication details see Supporting Information~\cite{SI}). 
The motion of the cantilever is detected using a fiber-based laser interferometer operating at wavelength of 1510~nm. 
MFM measurements are performed in a frequency detection mode \cite{Albrecht1991}, in which the cantilever is self-oscillated  at its resonant frequency, and the frequency is measured using phase-locked-loop (PLL). 
A PID feedback loop  is used to maintain a constant oscillation amplitude between 2.5-7.5~nm and to measure changes in dissipation. 
A silicon chip carrying the superconducting structures is mounted below the cantilever on a stack of nano-positioners, that permits approach, coarse positioning  and scanning of the superconducting device with respect to the cantilever. 
All scanning is done at a fixed height $h$ between 0.2-2~$\mu$m above the surface of the sample. All measurements were conducted in vacuum in a continuous flow $\mathrm{^3He}$ refrigerator operating at a base temperature of 350~mK.

In this work we study narrow-wall aluminum rings with radii between 0.5-2~$\mu$m and wall width $w$ between 100-200~nm. The rings are divided into two, three or four sectors  by radial crossbars. The devices were patterned by e-beam lithography of PMMA resist spun on a 500-nm thick $\mathrm{SiO_2}$ layer grown on a silicon substrate. After exposure and development, the PMMA was metalized by e-beam evaporation with a 5-nm thick Ti adhesion layer and a 45-nm thick Al device layer. The ring structures were obtained by lift-off of the PMMA.
We measured more than fifteen structures; here we report the representative results for four of them.

We start by considering vortex states supported by a structure with a ring geometry divided in two halves by a crossbar (Fig.~\ref{Fig2}b).
If the wall width is sufficiently small compared to the superconducting coherence length, which applies to all the structures reported in this work, then the vortex sates are described by pairs of winding numbers $(n_1, n_2)$ of the superconducting phase around the bottom and top halves of the ring.
For narrow and thin-wall structures, the free energy of the vortex state depends only on $(n_1, n_2)$ and fluxes $(\phi_1$, $\phi_2)$ threading the two halves of the structure, where $\phi$ is the magnetic flux in units of the flux quantum $\Phi_0=h/2e$.
For a symmetric ring with a crossbar (see Supplementary Information\cite{SI}),  the energy can be expressed as
\begin{equation}
F\propto \frac1{1+\beta}\left( \tilde{\phi_1}^2+2\beta\tilde{\phi_1}\tilde{\phi_2}+\tilde{\phi_2}^2
\right)
\label{Energy}
\end{equation}
where $\tilde{\phi_i}=\phi_i-n_i$.
The parameter $\beta$ controls the coupling between the two halves of the ring.
The limiting cases of $\beta=0$ and  $\beta=1$ correspond to either two isolated sectors or a vanishing crossbar, respectively.
While in general both the magnetic and kinetic inductances contribute to the coupling parameter $\beta$, structures for which $w t < \lambda^2$, where $t$ is the device thickness and $\lambda$ is the superconducting penetration depth, the kinetic contribution dominates. For such devices, which includes the one shown in Fig.~\ref{Fig2}b, a value of $\beta\approx 0.39$ is expected\cite{SI}.
The region of stability of each vortex state  $(n_1, n_2)$  in  $(\phi_1, \phi_2)$ coordinates has a shape of a  hexagon centered on  $(n_1, n_2)$ point (Fig.~\ref{Fig2}c).
The stability diagram  of the vortex states is similar to the honeycomb stability diagram of charging states in a double quantum dot~\cite{Wiel2002}. This similarity arises because both systems share the same form of the effective energy.
The coupling parameter $\beta$ controls the shape of the hexagons, and can be determined experimentally from the modulation depth of hexagons $\beta =l_2/l_1$, as shown in Fig.~\ref{Fig2}c.

The image of the cantilever frequency vs. tip position allows us to directly map the cantilever position to a particular vortex state.
Figure~\ref{Fig2}a shows an $\Phi_0$-MFM scan of Ring~1, which has a radius of $R=0.95$~$\mu$m and wall width $w=133$~nm, divided by a crossbar of the same width into two halves.
The scan is taken with tip-surface separation $h=0.8$~$\mu$m, at 0.915~K, close to the superconducting transition temperature $T_c \approx$0.95~K.  In this regime near $T_c$, the transitions between vortex states become reversible.
The dark colored contours, shown in Fig.~\ref{Fig2}a, indicate the shift of the resonance frequency of the cantilever , corresponding to transitions between two tip-induced vortex states -- the number of vortices changes by one in at least one of the sectors of the structure.
At these points, the energies of two or more vortex states are degenerate and the  field modulation $\delta B_z$ caused by the tip oscillations drives the vortex transitions. The transitions are accompanied by a switching of the supercurrents in the structure between two distinct configurations, which causes a strong back-action on the cantilever and gives rise to a shift of the resonance frequency~\cite{Polshyn_2017, SI}.

Every position of the tip $(x_{tip},y_{tip},h)$  relative to the superconducting structure corresponds to certain values of $(\phi_1, \phi_2)$ induced in the structure. Thus, each   $\mathrm{\Phi_0}$-MFM image is equivalent to a nonlinear mapping from $(x_{tip},y_{tip},h)$ to ($\phi_1$, $\phi_2$) coordinates.
The mapping is determined by the magnetic flux induced in each sector of the device for a given tip position.
We use $\mathrm{\Phi_0}$-MFM image shown in Fig.~\ref{Fig2}a and known dimensions of Ring~1 to reconstruct the mapping $(x_{tip},y_{tip},h)\to (\phi_1, \phi_2)$.
We start with a model of the magnetic tip and  calculate $(\phi_1, \phi_2)$ for every point of the scan  $(x_{tip},y_{tip},h)$.
Then, using equation~\eqref{Energy}, we  find the positions of the vortex transitions for a given tip model and parameter $\beta$.
Further, the model of the tip is tuned to obtain a good match between the simulated and observed vortex transitions, as shown in Fig.~\ref{Fig2}d (see Supporting Information for details).
Best fit is obtained with $\beta=0.40\pm 0.02$, which is in close agreement with the expected value of 0.39.

In Figs.~\ref{Fig2}d-e, we show the correspondence between  selected tip positions in $\Phi_0$-MFM image of Ring~1 and the points $(\phi_1, \phi_2)$ on the vortex configuration diagram. Establishing this correspondence identifies all vortex states, some of which are labeled in  Fig~\ref{Fig2}a. The shaded region in  Figs.~\ref{Fig2}e highlights the vortex states that were accessed during the scan.
It is worth noting, that in a homogeneous magnetic field, only points for which $\phi_1=\phi_2$ can be accessed.

The spatial distribution of the static magnetic field $B_z(x,y)$ and modulation $\delta B_z (x,y)$, obtained from matching vortex transitions in Fig.~\ref{Fig2}a,  are shown in Figs.~\ref{Fig2}f and \ref{Fig2}g respectively. The field modulation is highly spatially inhomogeneous and has two peaks with the opposite sign -- one in front and one behind the cantilever about $\sim 1\,\mathrm{\mu m}$ apart, in the oscillation direction of the cantilever.
This spatially separated character of the modulation in  $\Phi_0$-MFM enables efficiently driving transitions between different multi-vortex states.

Most of the static flux generated by the magnetic tip is confined to an area of size $\sim h^2$. Thus, it is possible to vary the total number of vortices in the structure by simply changing the tip-surface separation.
Figures~\ref{Fig3}a-d show a series of $\Phi_0$-MFM images of Ring~2 ($R=1.99$~$\mathrm{\mu m}$, $w=230\, \text{nm}$) taken  at several tip-surface separations $15$~mK below the superconducting transition $T_c=1.199$~K of the structure.
As $h$ decreases from 1.5~$\mathrm{\mu}$m to 0.8~$\mathrm{\mu}$m, the flux induced by the tip through the structure grows, and the maximum total number of vortices imaged in a scan increases from 7 to 13.
Even at the smallest tip-surface separation, the transitions between vortex states are extremely sharp ($\sim 15$~nm in real space) and manifest clear honeycomb pattern as shown in  the fine scan of the selected area of Fig.~\ref{Fig3}d, shown in Fig.~\ref{Fig3}e.

\begin{figure*}[t]
\centering
\includegraphics[]{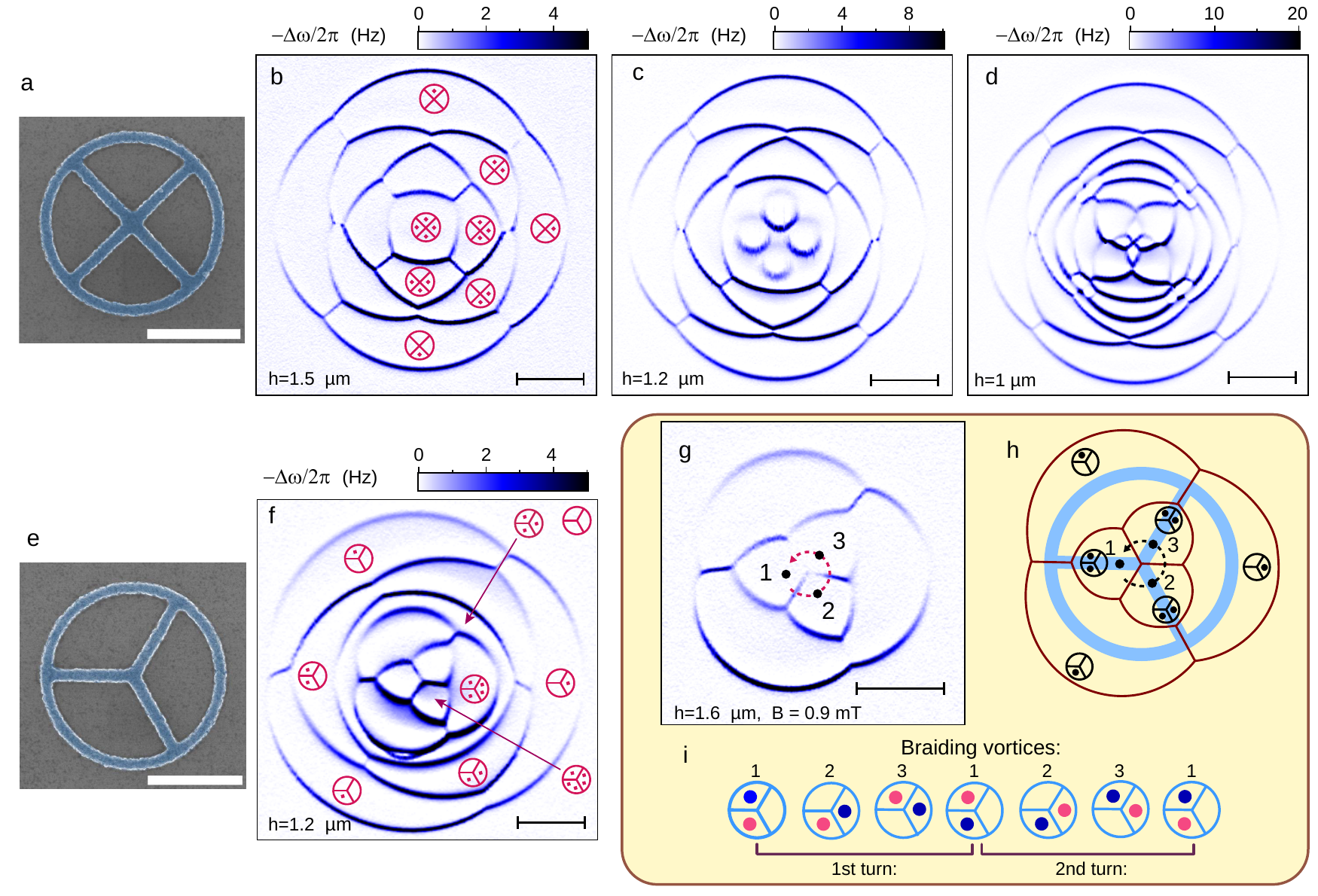}
\caption[] {Multi-vortex states in complex multi-loop structures. 
(a),(e) SEM images of Ring 3 and Ring 4. (b-d), (f) $\Phi_0$-MFM images for Ring~3 and Ring~4 measured at the tip height indicated on each panel. Selected vortex states are labeled with the diagrams of vortex configurations. 
(h) A schematic diagram showing the tip trajectory for performing a braiding operation for two vortices in the three-sector device (e). (g) Shows the $\Phi_0$-MFM image obtained under the experimental conditions necessary to perform braiding. The dashed arrow indicates the proposed trajectory of the magnetic tip with intermediate states ``1'', ``2'' and ``3''.
(i) Sequence of manipulations that braids two vortices. Two vortices are shown in red and blue only to emphasize their movements. The first turn of the loop swaps the vortices and the second completes the winding of one vortex around each other.
All scale bars are 1 $\mathrm{\mu m}$.}
\label{Fig4}
\end{figure*}

Controlling the dynamics of vortex transitions is crucial  for  realizing  metastable vortex states. $\Phi_0$-MFM provides a unique way of visualizing the vortex energy barrier between different parts of the device. We measure the vortex dynamics in our superconducting devices using a previously described technique, which relies on the dynamical interaction between thermally activated phase slips~(TAPS)~\cite{Little1967, Langer1967} in the superconducting device and the cantilever~\cite{Polshyn_2017}. Near $T_c$, the field modulation caused by the motion of the tip in the presence of TAPS can give rise to a dynamical force that significantly shifts the frequency and dissipation of the cantilever. In the regime where the vortex transition rate $\nu_r$ becomes comparable to the cantilever frequency $\omega_0$, a simple relationship exists between $\nu_r$ and the variations in the cantilever frequency $\Delta\omega$ and dissipation $\Delta\gamma$ ~\cite{Polshyn_2017,SI}.
\begin{equation}\label{eq:phaseSlipRate}
\frac{ \nu_r}{\omega_0}=\frac{ \Delta \omega}{\Delta \gamma}
\end{equation}

Figure~\ref{Fig3}f shows images of $\nu_r$ calculated using (\ref{eq:phaseSlipRate}), for the area marked in Fig.~\ref{Fig3}e. This region contains three individual vortex transitions: (A)$(5,5) \leftrightarrow (5,4)$, (B)$(5,5) \leftrightarrow (6,4)$ and (C)$(5,4)\leftrightarrow(6,4)$, measured at several temperatures near $T_c$.
The color range in Fig.~\ref{Fig3}f represents the measured relaxation rate $\nu_r/\omega_0 =\Delta\omega /\Delta \gamma$, while  the  brightness represents the magnitude of the signal $\sqrt{\Delta\omega^2+ \Delta \gamma^2}$.
As is evident from Fig.~\ref{Fig3}f, the three transition rates $\nu_r$ vary differently with respect to temperature, indicating slight differences in the energy barrier responsible for vortex entry corresponding to the different transitions (for more details see \cite{SI}). Transitions A and C correspond to vortices entering/leaving the bottom and top halves of the ring respectively, while  transition B corresponds to a vortex moving from one half of the ring to the other (see inset in Fig.~\ref{Fig3}e). For each type of transition, the corresponding phase slip occurs in a different part of the device. The variations in the transition rates could be result of small variations in the crossectional area of the device or the presence of defects.

We can generate more complex vortex states by fabricating superconducting structures with larger number of partitions. For the purposes of demonstration, we study vortex states hosted by circular structures partitioned into three and four equal area sectors, as shown in Fig.~\ref{Fig4}a and \ref{Fig4}e.
$\mathrm{\Phi_0}$-MFM images of the vortex states in these structures reveal complicated patters of transitions that evolve with tip-sample separation (Fig.~\ref{Fig4}b-d). Despite the complexity of the transition patterns however, the majority of vortex states can be easily identified from symmetry considerations, as shown in Fig.~\ref{Fig4}b and Fig.~\ref{Fig4}f.
In general, the full set of transitions observed in a given image can be calculated using a model for the field distribution from the tip and the analytic form of the free energy of the vortex states.
Images Fig.~\ref{Fig4}b-d,f demonstrate several unique capabilities of $\mathrm{\Phi_0}$-MFM approach:
1) The majority of the observed vortex states are stable only in the presence of the inhomogeneous magnetic field from the tip. Therefore, the ability to scan the magnetic particle provides a robust and highly reliable means of accessing a large range of vortex configurations. (2)~The $\Phi_0$-MFM images provide a deterministic mapping between the position of the cantilever and a particular vortex state. This mapping enables a simple and intuitive means of evolving the system through a desired trajectory of vortex configurations by simply tracing the tip through the corresponding trajectory in space.

As a starting point, we demonstrate a method for braiding two vortices in the three-sector device shown in Fig.~\ref{Fig4}e.
We choose $h$=1.6~$\mathrm{\mu}$m and a uniform external field of $B$=0.9~mT, such that for tip positions close to the center of the ring, only two of the sectors are populated with vortices (see Fig.~\ref{Fig4}g,h).
One complete circular motion of the tip around the center drives the structure through a sequence of vortex states, as shown in Fig.~\ref{Fig4}i, and exchanges the positions of the vortices.
Two full circles of the magnetic tip will accomplish a winding of one vortex around another.
While, in our system we control fluxons in a superconducting wire network, the same idea could be applied to Josephson or Abrikosov vortices in appropriate superconducting structures.

In summary, we have shown that $\Phi_0$-MFM can be used to probe complicated multi-vortex states in narrow-wall multiply connected superconducting structures. The spatially inhomogeneous magnetic field generated by MFM tip enables stabilization of complex vortex states that are not accessible in a uniform field. 
Strong interaction of the magnetic tip and supercurrents  in superconducting structure at points where tip-induced transitions occur enables the mapping of vortex transitions and the characterization of their dynamic properties. 
This approach provides a versatile way to induce, identify, and manipulate complex multi-vortex states in superconducting structures. 
Finally, as a potentially useful application of the technique, we describe how $\Phi_0$-MFM may be used to braid a pair of vortices around each other, which might be useful for experiments with Majorana bound states.
Unique capability to work with multi-vortex or complex vortex states opens new possibilities in the exploration of vortex interactions and designing new vortex-based superconducting devices. 

\section{acknowledgement}
 We are grateful to Nadya Mason, Taylor Hughes and Alexey Bezryadin for useful discussions. This work was supported by the DOE Basic Energy Sciences under DE-SC0012649, the Department of Physics and the Frederick Seitz Materials Research Laboratory Central Facilities at the University of Illinois.


\begin{thebibliography}{43}%
\makeatletter
\providecommand \@ifxundefined [1]{%
 \@ifx{#1\undefined}
}%
\providecommand \@ifnum [1]{%
 \ifnum #1\expandafter \@firstoftwo
 \else \expandafter \@secondoftwo
 \fi
}%
\providecommand \@ifx [1]{%
 \ifx #1\expandafter \@firstoftwo
 \else \expandafter \@secondoftwo
 \fi
}%
\providecommand \natexlab [1]{#1}%
\providecommand \enquote  [1]{``#1''}%
\providecommand \bibnamefont  [1]{#1}%
\providecommand \bibfnamefont [1]{#1}%
\providecommand \citenamefont [1]{#1}%
\providecommand \href@noop [0]{\@secondoftwo}%
\providecommand \href [0]{\begingroup \@sanitize@url \@href}%
\providecommand \@href[1]{\@@startlink{#1}\@@href}%
\providecommand \@@href[1]{\endgroup#1\@@endlink}%
\providecommand \@sanitize@url [0]{\catcode `\\12\catcode `\$12\catcode
  `\&12\catcode `\#12\catcode `\^12\catcode `\_12\catcode `\%12\relax}%
\providecommand \@@startlink[1]{}%
\providecommand \@@endlink[0]{}%
\providecommand \url  [0]{\begingroup\@sanitize@url \@url }%
\providecommand \@url [1]{\endgroup\@href {#1}{\urlprefix }}%
\providecommand \urlprefix  [0]{URL }%
\providecommand \Eprint [0]{\href }%
\providecommand \doibase [0]{http://dx.doi.org/}%
\providecommand \selectlanguage [0]{\@gobble}%
\providecommand \bibinfo  [0]{\@secondoftwo}%
\providecommand \bibfield  [0]{\@secondoftwo}%
\providecommand \translation [1]{[#1]}%
\providecommand \BibitemOpen [0]{}%
\providecommand \bibitemStop [0]{}%
\providecommand \bibitemNoStop [0]{.\EOS\space}%
\providecommand \EOS [0]{\spacefactor3000\relax}%
\providecommand \BibitemShut  [1]{\csname bibitem#1\endcsname}%
\let\auto@bib@innerbib\@empty
\bibitem [{\citenamefont {Tinkham}(1996)}]{Tinkham1996}%
  \BibitemOpen
  \bibfield  {author} {\bibinfo {author} {\bibfnamefont {M.}~\bibnamefont
  {Tinkham}},\ }\href@noop {} {\emph {\bibinfo {title} {Introduction to
  superconductivity}}}\ (\bibinfo  {publisher} {Dover Publications Inc.},\
  \bibinfo {year} {1996})\BibitemShut {NoStop}%
\bibitem [{\citenamefont {Bezryadin}\ \emph {et~al.}(1996)\citenamefont
  {Bezryadin}, \citenamefont {Ovchinnikov},\ and\ \citenamefont
  {Pannetier}}]{Bezryadin_nucleation_1996}%
  \BibitemOpen
  \bibfield  {author} {\bibinfo {author} {\bibfnamefont {A.}~\bibnamefont
  {Bezryadin}}, \bibinfo {author} {\bibfnamefont {Y.~N.}\ \bibnamefont
  {Ovchinnikov}}, \ and\ \bibinfo {author} {\bibfnamefont {B.}~\bibnamefont
  {Pannetier}},\ }\href {\doibase 10.1103/PhysRevB.53.8553} {\bibfield
  {journal} {\bibinfo  {journal} {Phys. Rev. B}\ }\textbf {\bibinfo {volume}
  {53}},\ \bibinfo {pages} {8553} (\bibinfo {year} {1996})}\BibitemShut
  {NoStop}%
\bibitem [{\citenamefont {Geim}\ \emph {et~al.}(1997)\citenamefont {Geim},
  \citenamefont {Grigorieva}, \citenamefont {Dubonos}, \citenamefont {Lok},
  \citenamefont {Maan}, \citenamefont {Filippov},\ and\ \citenamefont
  {Peeters}}]{Geim1997}%
  \BibitemOpen
  \bibfield  {author} {\bibinfo {author} {\bibfnamefont {A.}~\bibnamefont
  {Geim}}, \bibinfo {author} {\bibfnamefont {I.}~\bibnamefont {Grigorieva}},
  \bibinfo {author} {\bibfnamefont {S.}~\bibnamefont {Dubonos}}, \bibinfo
  {author} {\bibfnamefont {J.}~\bibnamefont {Lok}}, \bibinfo {author}
  {\bibfnamefont {J.}~\bibnamefont {Maan}}, \bibinfo {author} {\bibfnamefont
  {A.}~\bibnamefont {Filippov}}, \ and\ \bibinfo {author} {\bibfnamefont
  {F.}~\bibnamefont {Peeters}},\ }\href {\doibase 10.1038/36797} {\bibfield
  {journal} {\bibinfo  {journal} {Nature}\ }\textbf {\bibinfo {volume} {390}},\
  \bibinfo {pages} {259} (\bibinfo {year} {1997})}\BibitemShut {NoStop}%
\bibitem [{\citenamefont {Geim}\ \emph {et~al.}(1998)\citenamefont {Geim},
  \citenamefont {Dubonos}, \citenamefont {Lok}, \citenamefont {Henini},\ and\
  \citenamefont {Maan}}]{Geim1998}%
  \BibitemOpen
  \bibfield  {author} {\bibinfo {author} {\bibfnamefont {A.}~\bibnamefont
  {Geim}}, \bibinfo {author} {\bibfnamefont {S.}~\bibnamefont {Dubonos}},
  \bibinfo {author} {\bibfnamefont {J.}~\bibnamefont {Lok}}, \bibinfo {author}
  {\bibfnamefont {M.}~\bibnamefont {Henini}}, \ and\ \bibinfo {author}
  {\bibfnamefont {J.}~\bibnamefont {Maan}},\ }\href {\doibase 10.1038/24110}
  {\bibfield  {journal} {\bibinfo  {journal} {Nature}\ }\textbf {\bibinfo
  {volume} {396}},\ \bibinfo {pages} {144} (\bibinfo {year}
  {1998})}\BibitemShut {NoStop}%
\bibitem [{\citenamefont {Morelle}\ \emph {et~al.}(2004)\citenamefont
  {Morelle}, \citenamefont {Golubovi\ifmmode~\acute{c}\else \'{c}\fi{}},\ and\
  \citenamefont {Moshchalkov}}]{Morelle2004}%
  \BibitemOpen
  \bibfield  {author} {\bibinfo {author} {\bibfnamefont {M.}~\bibnamefont
  {Morelle}}, \bibinfo {author} {\bibfnamefont {D.~c. v.~S.}\ \bibnamefont
  {Golubovi\ifmmode~\acute{c}\else \'{c}\fi{}}}, \ and\ \bibinfo {author}
  {\bibfnamefont {V.~V.}\ \bibnamefont {Moshchalkov}},\ }\href {\doibase
  10.1103/PhysRevB.70.144528} {\bibfield  {journal} {\bibinfo  {journal} {Phys.
  Rev. B}\ }\textbf {\bibinfo {volume} {70}},\ \bibinfo {pages} {144528}
  (\bibinfo {year} {2004})}\BibitemShut {NoStop}%
\bibitem [{\citenamefont {Grigorieva}\ \emph {et~al.}(2006)\citenamefont
  {Grigorieva}, \citenamefont {Escoffier}, \citenamefont {Richardson},
  \citenamefont {Vinnikov}, \citenamefont {Dubonos},\ and\ \citenamefont
  {Oboznov}}]{Grigorieva2006}%
  \BibitemOpen
  \bibfield  {author} {\bibinfo {author} {\bibfnamefont {I.~V.}\ \bibnamefont
  {Grigorieva}}, \bibinfo {author} {\bibfnamefont {W.}~\bibnamefont
  {Escoffier}}, \bibinfo {author} {\bibfnamefont {J.}~\bibnamefont
  {Richardson}}, \bibinfo {author} {\bibfnamefont {L.~Y.}\ \bibnamefont
  {Vinnikov}}, \bibinfo {author} {\bibfnamefont {S.}~\bibnamefont {Dubonos}}, \
  and\ \bibinfo {author} {\bibfnamefont {V.}~\bibnamefont {Oboznov}},\ }\href
  {\doibase 10.1103/PhysRevLett.96.077005} {\bibfield  {journal} {\bibinfo
  {journal} {Phys. Rev. Lett.}\ }\textbf {\bibinfo {volume} {96}},\ \bibinfo
  {pages} {077005} (\bibinfo {year} {2006})}\BibitemShut {NoStop}%
\bibitem [{\citenamefont {Timmermans}\ \emph {et~al.}(2016)\citenamefont
  {Timmermans}, \citenamefont {Serrier-Garcia}, \citenamefont {Perini},
  \citenamefont {Van~de Vondel},\ and\ \citenamefont
  {Moshchalkov}}]{Timmermans_direct_2016}%
  \BibitemOpen
  \bibfield  {author} {\bibinfo {author} {\bibfnamefont {M.}~\bibnamefont
  {Timmermans}}, \bibinfo {author} {\bibfnamefont {L.}~\bibnamefont
  {Serrier-Garcia}}, \bibinfo {author} {\bibfnamefont {M.}~\bibnamefont
  {Perini}}, \bibinfo {author} {\bibfnamefont {J.}~\bibnamefont {Van~de
  Vondel}}, \ and\ \bibinfo {author} {\bibfnamefont {V.~V.}\ \bibnamefont
  {Moshchalkov}},\ }\href {\doibase 10.1103/PhysRevB.93.054514} {\bibfield
  {journal} {\bibinfo  {journal} {Phys. Rev. B}\ }\textbf {\bibinfo {volume}
  {93}},\ \bibinfo {pages} {054514} (\bibinfo {year} {2016})}\BibitemShut
  {NoStop}%
\bibitem [{\citenamefont {Embon}\ \emph {et~al.}(2017)\citenamefont {Embon},
  \citenamefont {Anahory}, \citenamefont {Jeli{\'c}}, \citenamefont {Lachman},
  \citenamefont {Myasoedov}, \citenamefont {Huber}, \citenamefont {Mikitik},
  \citenamefont {Silhanek}, \citenamefont {Milo{\v{s}}evi{\'c}}, \citenamefont
  {Gurevich} \emph {et~al.}}]{embon2017imaging}%
  \BibitemOpen
  \bibfield  {author} {\bibinfo {author} {\bibfnamefont {L.}~\bibnamefont
  {Embon}}, \bibinfo {author} {\bibfnamefont {Y.}~\bibnamefont {Anahory}},
  \bibinfo {author} {\bibfnamefont {{\v{Z}}.~L.}\ \bibnamefont {Jeli{\'c}}},
  \bibinfo {author} {\bibfnamefont {E.~O.}\ \bibnamefont {Lachman}}, \bibinfo
  {author} {\bibfnamefont {Y.}~\bibnamefont {Myasoedov}}, \bibinfo {author}
  {\bibfnamefont {M.~E.}\ \bibnamefont {Huber}}, \bibinfo {author}
  {\bibfnamefont {G.~P.}\ \bibnamefont {Mikitik}}, \bibinfo {author}
  {\bibfnamefont {A.~V.}\ \bibnamefont {Silhanek}}, \bibinfo {author}
  {\bibfnamefont {M.~V.}\ \bibnamefont {Milo{\v{s}}evi{\'c}}}, \bibinfo
  {author} {\bibfnamefont {A.}~\bibnamefont {Gurevich}},  \emph {et~al.},\
  }\href {\doibase 10.1038/s41467-017-00089-3} {\bibfield  {journal} {\bibinfo
  {journal} {Nature communications}\ }\textbf {\bibinfo {volume} {8}},\
  \bibinfo {pages} {85} (\bibinfo {year} {2017})}\BibitemShut {NoStop}%
\bibitem [{\citenamefont {Peeters}\ \emph {et~al.}(2000)\citenamefont
  {Peeters}, \citenamefont {Schweigert}, \citenamefont {Baelus},\ and\
  \citenamefont {Deo}}]{Peeters2000}%
  \BibitemOpen
  \bibfield  {author} {\bibinfo {author} {\bibfnamefont {F.}~\bibnamefont
  {Peeters}}, \bibinfo {author} {\bibfnamefont {V.}~\bibnamefont {Schweigert}},
  \bibinfo {author} {\bibfnamefont {B.}~\bibnamefont {Baelus}}, \ and\ \bibinfo
  {author} {\bibfnamefont {P.}~\bibnamefont {Deo}},\ }\href {\doibase
  http://dx.doi.org/10.1016/S0921-4534(99)00681-4} {\bibfield  {journal}
  {\bibinfo  {journal} {Physica C: Superconductivity}\ }\textbf {\bibinfo
  {volume} {332}},\ \bibinfo {pages} {255 } (\bibinfo {year}
  {2000})}\BibitemShut {NoStop}%
\bibitem [{\citenamefont {Baelus}\ \emph {et~al.}(2000)\citenamefont {Baelus},
  \citenamefont {Peeters},\ and\ \citenamefont {Schweigert}}]{Baelus2000}%
  \BibitemOpen
  \bibfield  {author} {\bibinfo {author} {\bibfnamefont {B.~J.}\ \bibnamefont
  {Baelus}}, \bibinfo {author} {\bibfnamefont {F.~M.}\ \bibnamefont {Peeters}},
  \ and\ \bibinfo {author} {\bibfnamefont {V.~A.}\ \bibnamefont {Schweigert}},\
  }\href {\doibase 10.1103/PhysRevB.61.9734} {\bibfield  {journal} {\bibinfo
  {journal} {Phys. Rev. B}\ }\textbf {\bibinfo {volume} {61}},\ \bibinfo
  {pages} {9734} (\bibinfo {year} {2000})}\BibitemShut {NoStop}%
\bibitem [{\citenamefont {Baelus}\ \emph {et~al.}(2001)\citenamefont {Baelus},
  \citenamefont {Peeters},\ and\ \citenamefont {Schweigert}}]{Baelus_2001}%
  \BibitemOpen
  \bibfield  {author} {\bibinfo {author} {\bibfnamefont {B.~J.}\ \bibnamefont
  {Baelus}}, \bibinfo {author} {\bibfnamefont {F.~M.}\ \bibnamefont {Peeters}},
  \ and\ \bibinfo {author} {\bibfnamefont {V.~A.}\ \bibnamefont {Schweigert}},\
  }\href {\doibase 10.1103/PhysRevB.63.144517} {\bibfield  {journal} {\bibinfo
  {journal} {Phys. Rev. B}\ }\textbf {\bibinfo {volume} {63}},\ \bibinfo
  {pages} {144517} (\bibinfo {year} {2001})}\BibitemShut {NoStop}%
\bibitem [{\citenamefont {Baelus}\ and\ \citenamefont
  {Peeters}(2002)}]{Baelus_2002}%
  \BibitemOpen
  \bibfield  {author} {\bibinfo {author} {\bibfnamefont {B.~J.}\ \bibnamefont
  {Baelus}}\ and\ \bibinfo {author} {\bibfnamefont {F.~M.}\ \bibnamefont
  {Peeters}},\ }\href {\doibase 10.1103/PhysRevB.65.104515} {\bibfield
  {journal} {\bibinfo  {journal} {Phys. Rev. B}\ }\textbf {\bibinfo {volume}
  {65}},\ \bibinfo {pages} {104515} (\bibinfo {year} {2002})}\BibitemShut
  {NoStop}%
\bibitem [{\citenamefont {Aharonov}\ and\ \citenamefont
  {Casher}(1984)}]{Aharonov1984}%
  \BibitemOpen
  \bibfield  {author} {\bibinfo {author} {\bibfnamefont {Y.}~\bibnamefont
  {Aharonov}}\ and\ \bibinfo {author} {\bibfnamefont {A.}~\bibnamefont
  {Casher}},\ }\href {\doibase 10.1103/PhysRevLett.53.319} {\bibfield
  {journal} {\bibinfo  {journal} {Phys. Rev. Lett.}\ }\textbf {\bibinfo
  {volume} {53}},\ \bibinfo {pages} {319} (\bibinfo {year} {1984})}\BibitemShut
  {NoStop}%
\bibitem [{\citenamefont {Kosterlitz}\ and\ \citenamefont
  {Thouless}(1973)}]{Kosterlitz_1973}%
  \BibitemOpen
  \bibfield  {author} {\bibinfo {author} {\bibfnamefont {J.~M.}\ \bibnamefont
  {Kosterlitz}}\ and\ \bibinfo {author} {\bibfnamefont {D.~J.}\ \bibnamefont
  {Thouless}},\ }\href {\doibase 10.1088/0022-3719/6/7/010} {\bibfield
  {journal} {\bibinfo  {journal} {Journal of Physics C: Solid State Physics}\
  }\textbf {\bibinfo {volume} {6}},\ \bibinfo {pages} {1181} (\bibinfo {year}
  {1973})}\BibitemShut {NoStop}%
\bibitem [{\citenamefont {Fu}\ and\ \citenamefont {Kane}(2008)}]{Fu2008}%
  \BibitemOpen
  \bibfield  {author} {\bibinfo {author} {\bibfnamefont {L.}~\bibnamefont
  {Fu}}\ and\ \bibinfo {author} {\bibfnamefont {C.~L.}\ \bibnamefont {Kane}},\
  }\href {\doibase 10.1103/PhysRevLett.100.096407} {\bibfield  {journal}
  {\bibinfo  {journal} {Phys. Rev. Lett.}\ }\textbf {\bibinfo {volume} {100}},\
  \bibinfo {pages} {096407} (\bibinfo {year} {2008})}\BibitemShut {NoStop}%
\bibitem [{\citenamefont {Beenakker}(2013)}]{beenaker_2013}%
  \BibitemOpen
  \bibfield  {author} {\bibinfo {author} {\bibfnamefont {C.}~\bibnamefont
  {Beenakker}},\ }\href {\doibase 10.1146/annurev-conmatphys-030212-184337}
  {\bibfield  {journal} {\bibinfo  {journal} {Annual Review of Condensed Matter
  Physics}\ }\textbf {\bibinfo {volume} {4}},\ \bibinfo {pages} {113} (\bibinfo
  {year} {2013})}\BibitemShut {NoStop}%
\bibitem [{\citenamefont {Ren}\ \emph {et~al.}(2019)\citenamefont {Ren},
  \citenamefont {Pientka}, \citenamefont {Hart}, \citenamefont {Pierce},
  \citenamefont {Kosowsky}, \citenamefont {Lunczer}, \citenamefont {Schlereth},
  \citenamefont {Scharf}, \citenamefont {Hankiewicz}, \citenamefont
  {Molenkamp}, \citenamefont {Halperin},\ and\ \citenamefont
  {Yacoby}}]{Ren_2019}%
  \BibitemOpen
  \bibfield  {author} {\bibinfo {author} {\bibfnamefont {H.}~\bibnamefont
  {Ren}}, \bibinfo {author} {\bibfnamefont {F.}~\bibnamefont {Pientka}},
  \bibinfo {author} {\bibfnamefont {S.}~\bibnamefont {Hart}}, \bibinfo {author}
  {\bibfnamefont {A.~T.}\ \bibnamefont {Pierce}}, \bibinfo {author}
  {\bibfnamefont {M.}~\bibnamefont {Kosowsky}}, \bibinfo {author}
  {\bibfnamefont {L.}~\bibnamefont {Lunczer}}, \bibinfo {author} {\bibfnamefont
  {R.}~\bibnamefont {Schlereth}}, \bibinfo {author} {\bibfnamefont
  {B.}~\bibnamefont {Scharf}}, \bibinfo {author} {\bibfnamefont {E.~M.}\
  \bibnamefont {Hankiewicz}}, \bibinfo {author} {\bibfnamefont {L.~W.}\
  \bibnamefont {Molenkamp}}, \bibinfo {author} {\bibfnamefont {B.~I.}\
  \bibnamefont {Halperin}}, \ and\ \bibinfo {author} {\bibfnamefont
  {A.}~\bibnamefont {Yacoby}},\ }\href {\doibase 10.1038/s41586-019-1148-9}
  {\bibfield  {journal} {\bibinfo  {journal} {Nature}\ } (\bibinfo {year}
  {2019}),\ 10.1038/s41586-019-1148-9}\BibitemShut {NoStop}%
\bibitem [{\citenamefont {Zhang}\ \emph {et~al.}(2019)\citenamefont {Zhang},
  \citenamefont {Wang}, \citenamefont {Wu}, \citenamefont {Yaji}, \citenamefont
  {Ishida}, \citenamefont {Kohama}, \citenamefont {Dai}, \citenamefont {Sun},
  \citenamefont {Bareille}, \citenamefont {Kuroda}, \citenamefont {Kondo},
  \citenamefont {Okazaki}, \citenamefont {Kindo}, \citenamefont {Wang},
  \citenamefont {Jin}, \citenamefont {Hu}, \citenamefont {Thomale},
  \citenamefont {Sumida}, \citenamefont {Wu}, \citenamefont {Miyamoto},
  \citenamefont {Okuda}, \citenamefont {Ding}, \citenamefont {Gu},
  \citenamefont {Tamegai}, \citenamefont {Kawakami}, \citenamefont {Sato},\
  and\ \citenamefont {Shin}}]{zhang_multiple_2019}%
  \BibitemOpen
  \bibfield  {author} {\bibinfo {author} {\bibfnamefont {P.}~\bibnamefont
  {Zhang}}, \bibinfo {author} {\bibfnamefont {Z.}~\bibnamefont {Wang}},
  \bibinfo {author} {\bibfnamefont {X.}~\bibnamefont {Wu}}, \bibinfo {author}
  {\bibfnamefont {K.}~\bibnamefont {Yaji}}, \bibinfo {author} {\bibfnamefont
  {Y.}~\bibnamefont {Ishida}}, \bibinfo {author} {\bibfnamefont
  {Y.}~\bibnamefont {Kohama}}, \bibinfo {author} {\bibfnamefont
  {G.}~\bibnamefont {Dai}}, \bibinfo {author} {\bibfnamefont {Y.}~\bibnamefont
  {Sun}}, \bibinfo {author} {\bibfnamefont {C.}~\bibnamefont {Bareille}},
  \bibinfo {author} {\bibfnamefont {K.}~\bibnamefont {Kuroda}}, \bibinfo
  {author} {\bibfnamefont {T.}~\bibnamefont {Kondo}}, \bibinfo {author}
  {\bibfnamefont {K.}~\bibnamefont {Okazaki}}, \bibinfo {author} {\bibfnamefont
  {K.}~\bibnamefont {Kindo}}, \bibinfo {author} {\bibfnamefont
  {X.}~\bibnamefont {Wang}}, \bibinfo {author} {\bibfnamefont {C.}~\bibnamefont
  {Jin}}, \bibinfo {author} {\bibfnamefont {J.}~\bibnamefont {Hu}}, \bibinfo
  {author} {\bibfnamefont {R.}~\bibnamefont {Thomale}}, \bibinfo {author}
  {\bibfnamefont {K.}~\bibnamefont {Sumida}}, \bibinfo {author} {\bibfnamefont
  {S.}~\bibnamefont {Wu}}, \bibinfo {author} {\bibfnamefont {K.}~\bibnamefont
  {Miyamoto}}, \bibinfo {author} {\bibfnamefont {T.}~\bibnamefont {Okuda}},
  \bibinfo {author} {\bibfnamefont {H.}~\bibnamefont {Ding}}, \bibinfo {author}
  {\bibfnamefont {G.~D.}\ \bibnamefont {Gu}}, \bibinfo {author} {\bibfnamefont
  {T.}~\bibnamefont {Tamegai}}, \bibinfo {author} {\bibfnamefont
  {T.}~\bibnamefont {Kawakami}}, \bibinfo {author} {\bibfnamefont
  {M.}~\bibnamefont {Sato}}, \ and\ \bibinfo {author} {\bibfnamefont
  {S.}~\bibnamefont {Shin}},\ }\href {\doibase 10.1038/s41567-018-0280-z}
  {\bibfield  {journal} {\bibinfo  {journal} {Nature Physics}\ }\textbf
  {\bibinfo {volume} {15}},\ \bibinfo {pages} {41} (\bibinfo {year}
  {2019})}\BibitemShut {NoStop}%
\bibitem [{\citenamefont {Wang}\ \emph {et~al.}(2018)\citenamefont {Wang},
  \citenamefont {Kong}, \citenamefont {Fan}, \citenamefont {Chen},
  \citenamefont {Zhu}, \citenamefont {Liu}, \citenamefont {Cao}, \citenamefont
  {Sun}, \citenamefont {Du}, \citenamefont {Schneeloch}, \citenamefont {Zhong},
  \citenamefont {Gu}, \citenamefont {Fu}, \citenamefont {Ding},\ and\
  \citenamefont {Gao}}]{wang_evidence_2018}%
  \BibitemOpen
  \bibfield  {author} {\bibinfo {author} {\bibfnamefont {D.}~\bibnamefont
  {Wang}}, \bibinfo {author} {\bibfnamefont {L.}~\bibnamefont {Kong}}, \bibinfo
  {author} {\bibfnamefont {P.}~\bibnamefont {Fan}}, \bibinfo {author}
  {\bibfnamefont {H.}~\bibnamefont {Chen}}, \bibinfo {author} {\bibfnamefont
  {S.}~\bibnamefont {Zhu}}, \bibinfo {author} {\bibfnamefont {W.}~\bibnamefont
  {Liu}}, \bibinfo {author} {\bibfnamefont {L.}~\bibnamefont {Cao}}, \bibinfo
  {author} {\bibfnamefont {Y.}~\bibnamefont {Sun}}, \bibinfo {author}
  {\bibfnamefont {S.}~\bibnamefont {Du}}, \bibinfo {author} {\bibfnamefont
  {J.}~\bibnamefont {Schneeloch}}, \bibinfo {author} {\bibfnamefont
  {R.}~\bibnamefont {Zhong}}, \bibinfo {author} {\bibfnamefont
  {G.}~\bibnamefont {Gu}}, \bibinfo {author} {\bibfnamefont {L.}~\bibnamefont
  {Fu}}, \bibinfo {author} {\bibfnamefont {H.}~\bibnamefont {Ding}}, \ and\
  \bibinfo {author} {\bibfnamefont {H.-J.}\ \bibnamefont {Gao}},\ }\href
  {\doibase 10.1126/science.aao1797} {\bibfield  {journal} {\bibinfo  {journal}
  {Science}\ }\textbf {\bibinfo {volume} {362}},\ \bibinfo {pages} {333}
  (\bibinfo {year} {2018})}\BibitemShut {NoStop}%
\bibitem [{\citenamefont {Liu}\ \emph {et~al.}(2018)\citenamefont {Liu},
  \citenamefont {Chen}, \citenamefont {Zhang}, \citenamefont {Peng},
  \citenamefont {Yan}, \citenamefont {Wen}, \citenamefont {Lou}, \citenamefont
  {Huang}, \citenamefont {Tian}, \citenamefont {Dong}, \citenamefont {Wang},
  \citenamefont {Bao}, \citenamefont {Wang}, \citenamefont {Yin}, \citenamefont
  {Zhao},\ and\ \citenamefont {Feng}}]{Liu_robust_2018}%
  \BibitemOpen
  \bibfield  {author} {\bibinfo {author} {\bibfnamefont {Q.}~\bibnamefont
  {Liu}}, \bibinfo {author} {\bibfnamefont {C.}~\bibnamefont {Chen}}, \bibinfo
  {author} {\bibfnamefont {T.}~\bibnamefont {Zhang}}, \bibinfo {author}
  {\bibfnamefont {R.}~\bibnamefont {Peng}}, \bibinfo {author} {\bibfnamefont
  {Y.-J.}\ \bibnamefont {Yan}}, \bibinfo {author} {\bibfnamefont {C.-H.-P.}\
  \bibnamefont {Wen}}, \bibinfo {author} {\bibfnamefont {X.}~\bibnamefont
  {Lou}}, \bibinfo {author} {\bibfnamefont {Y.-L.}\ \bibnamefont {Huang}},
  \bibinfo {author} {\bibfnamefont {J.-P.}\ \bibnamefont {Tian}}, \bibinfo
  {author} {\bibfnamefont {X.-L.}\ \bibnamefont {Dong}}, \bibinfo {author}
  {\bibfnamefont {G.-W.}\ \bibnamefont {Wang}}, \bibinfo {author}
  {\bibfnamefont {W.-C.}\ \bibnamefont {Bao}}, \bibinfo {author} {\bibfnamefont
  {Q.-H.}\ \bibnamefont {Wang}}, \bibinfo {author} {\bibfnamefont {Z.-P.}\
  \bibnamefont {Yin}}, \bibinfo {author} {\bibfnamefont {Z.-X.}\ \bibnamefont
  {Zhao}}, \ and\ \bibinfo {author} {\bibfnamefont {D.-L.}\ \bibnamefont
  {Feng}},\ }\href {\doibase 10.1103/PhysRevX.8.041056} {\bibfield  {journal}
  {\bibinfo  {journal} {Phys. Rev. X}\ }\textbf {\bibinfo {volume} {8}},\
  \bibinfo {pages} {041056} (\bibinfo {year} {2018})}\BibitemShut {NoStop}%
\bibitem [{\citenamefont {Zhang}\ \emph {et~al.}(2018)\citenamefont {Zhang},
  \citenamefont {Yaji}, \citenamefont {Hashimoto}, \citenamefont {Ota},
  \citenamefont {Kondo}, \citenamefont {Okazaki}, \citenamefont {Wang},
  \citenamefont {Wen}, \citenamefont {Gu}, \citenamefont {Ding},\ and\
  \citenamefont {Shin}}]{zhang_observation_2018}%
  \BibitemOpen
  \bibfield  {author} {\bibinfo {author} {\bibfnamefont {P.}~\bibnamefont
  {Zhang}}, \bibinfo {author} {\bibfnamefont {K.}~\bibnamefont {Yaji}},
  \bibinfo {author} {\bibfnamefont {T.}~\bibnamefont {Hashimoto}}, \bibinfo
  {author} {\bibfnamefont {Y.}~\bibnamefont {Ota}}, \bibinfo {author}
  {\bibfnamefont {T.}~\bibnamefont {Kondo}}, \bibinfo {author} {\bibfnamefont
  {K.}~\bibnamefont {Okazaki}}, \bibinfo {author} {\bibfnamefont
  {Z.}~\bibnamefont {Wang}}, \bibinfo {author} {\bibfnamefont {J.}~\bibnamefont
  {Wen}}, \bibinfo {author} {\bibfnamefont {G.~D.}\ \bibnamefont {Gu}},
  \bibinfo {author} {\bibfnamefont {H.}~\bibnamefont {Ding}}, \ and\ \bibinfo
  {author} {\bibfnamefont {S.}~\bibnamefont {Shin}},\ }\href {\doibase
  10.1126/science.aan4596} {\bibfield  {journal} {\bibinfo  {journal}
  {Science}\ }\textbf {\bibinfo {volume} {360}},\ \bibinfo {pages} {182}
  (\bibinfo {year} {2018})}\BibitemShut {NoStop}%
\bibitem [{\citenamefont {Kalisky}\ \emph {et~al.}(2009)\citenamefont
  {Kalisky}, \citenamefont {Kirtley}, \citenamefont {Nowadnick}, \citenamefont
  {Dinner}, \citenamefont {Zeldov}, \citenamefont {Ariando}, \citenamefont
  {Wenderich}, \citenamefont {Hilgenkamp}, \citenamefont {Feldmann},\ and\
  \citenamefont {Moler}}]{Kalisky2009}%
  \BibitemOpen
  \bibfield  {author} {\bibinfo {author} {\bibfnamefont {B.}~\bibnamefont
  {Kalisky}}, \bibinfo {author} {\bibfnamefont {J.~R.}\ \bibnamefont
  {Kirtley}}, \bibinfo {author} {\bibfnamefont {E.~A.}\ \bibnamefont
  {Nowadnick}}, \bibinfo {author} {\bibfnamefont {R.~B.}\ \bibnamefont
  {Dinner}}, \bibinfo {author} {\bibfnamefont {E.}~\bibnamefont {Zeldov}},
  \bibinfo {author} {\bibnamefont {Ariando}}, \bibinfo {author} {\bibfnamefont
  {S.}~\bibnamefont {Wenderich}}, \bibinfo {author} {\bibfnamefont
  {H.}~\bibnamefont {Hilgenkamp}}, \bibinfo {author} {\bibfnamefont {D.~M.}\
  \bibnamefont {Feldmann}}, \ and\ \bibinfo {author} {\bibfnamefont {K.~A.}\
  \bibnamefont {Moler}},\ }\href {\doibase 10.1063/1.3137164} {\bibfield
  {journal} {\bibinfo  {journal} {Applied Physics Letters}\ }\textbf {\bibinfo
  {volume} {94}},\ \bibinfo {pages} {202504} (\bibinfo {year}
  {2009})}\BibitemShut {NoStop}%
\bibitem [{\citenamefont {Embon}\ \emph {et~al.}(2015)\citenamefont {Embon},
  \citenamefont {Anahory}, \citenamefont {Suhov}, \citenamefont {Halbertal},
  \citenamefont {Cuppens}, \citenamefont {Yakovenko}, \citenamefont {Uri},
  \citenamefont {Myasoedov}, \citenamefont {Rappaport}, \citenamefont {Huber},
  \citenamefont {Gurevich},\ and\ \citenamefont {Zeldov}}]{Embon2015}%
  \BibitemOpen
  \bibfield  {author} {\bibinfo {author} {\bibfnamefont {L.}~\bibnamefont
  {Embon}}, \bibinfo {author} {\bibfnamefont {Y.}~\bibnamefont {Anahory}},
  \bibinfo {author} {\bibfnamefont {A.}~\bibnamefont {Suhov}}, \bibinfo
  {author} {\bibfnamefont {D.}~\bibnamefont {Halbertal}}, \bibinfo {author}
  {\bibfnamefont {J.}~\bibnamefont {Cuppens}}, \bibinfo {author} {\bibfnamefont
  {A.}~\bibnamefont {Yakovenko}}, \bibinfo {author} {\bibfnamefont
  {A.}~\bibnamefont {Uri}}, \bibinfo {author} {\bibfnamefont {Y.}~\bibnamefont
  {Myasoedov}}, \bibinfo {author} {\bibfnamefont {M.~L.}\ \bibnamefont
  {Rappaport}}, \bibinfo {author} {\bibfnamefont {M.~E.}\ \bibnamefont
  {Huber}}, \bibinfo {author} {\bibfnamefont {A.}~\bibnamefont {Gurevich}}, \
  and\ \bibinfo {author} {\bibfnamefont {E.}~\bibnamefont {Zeldov}},\ }\href
  {\doibase 10.1038/srep07598} {\bibfield  {journal} {\bibinfo  {journal}
  {Scientific Reports}\ }\textbf {\bibinfo {volume} {5}},\ \bibinfo {pages}
  {7598} (\bibinfo {year} {2015})}\BibitemShut {NoStop}%
\bibitem [{\citenamefont {Kalcheim}\ \emph {et~al.}(2017)\citenamefont
  {Kalcheim}, \citenamefont {Katzir}, \citenamefont {Zeides}, \citenamefont
  {Katz}, \citenamefont {Paltiel},\ and\ \citenamefont
  {Millo}}]{kalcheim_dynamic_2017}%
  \BibitemOpen
  \bibfield  {author} {\bibinfo {author} {\bibfnamefont {Y.}~\bibnamefont
  {Kalcheim}}, \bibinfo {author} {\bibfnamefont {E.}~\bibnamefont {Katzir}},
  \bibinfo {author} {\bibfnamefont {F.}~\bibnamefont {Zeides}}, \bibinfo
  {author} {\bibfnamefont {N.}~\bibnamefont {Katz}}, \bibinfo {author}
  {\bibfnamefont {Y.}~\bibnamefont {Paltiel}}, \ and\ \bibinfo {author}
  {\bibfnamefont {O.}~\bibnamefont {Millo}},\ }\href {\doibase
  10.1021/acs.nanolett.7b00179} {\bibfield  {journal} {\bibinfo  {journal}
  {Nano Letters}\ }\textbf {\bibinfo {volume} {17}},\ \bibinfo {pages} {2934}
  (\bibinfo {year} {2017})}\BibitemShut {NoStop}%
\bibitem [{\citenamefont {Ji}\ \emph {et~al.}(2016)\citenamefont {Ji},
  \citenamefont {Yuan}, \citenamefont {He}, \citenamefont {Jin}, \citenamefont
  {Zhu}, \citenamefont {Kong}, \citenamefont {Jia}, \citenamefont {Kang},
  \citenamefont {Jin},\ and\ \citenamefont {Wu}}]{ji_vortex_2016}%
  \BibitemOpen
  \bibfield  {author} {\bibinfo {author} {\bibfnamefont {J.}~\bibnamefont
  {Ji}}, \bibinfo {author} {\bibfnamefont {J.}~\bibnamefont {Yuan}}, \bibinfo
  {author} {\bibfnamefont {G.}~\bibnamefont {He}}, \bibinfo {author}
  {\bibfnamefont {B.}~\bibnamefont {Jin}}, \bibinfo {author} {\bibfnamefont
  {B.}~\bibnamefont {Zhu}}, \bibinfo {author} {\bibfnamefont {X.}~\bibnamefont
  {Kong}}, \bibinfo {author} {\bibfnamefont {X.}~\bibnamefont {Jia}}, \bibinfo
  {author} {\bibfnamefont {L.}~\bibnamefont {Kang}}, \bibinfo {author}
  {\bibfnamefont {K.}~\bibnamefont {Jin}}, \ and\ \bibinfo {author}
  {\bibfnamefont {P.}~\bibnamefont {Wu}},\ }\href {\doibase 10.1063/1.4971835}
  {\bibfield  {journal} {\bibinfo  {journal} {Applied Physics Letters}\
  }\textbf {\bibinfo {volume} {109}},\ \bibinfo {pages} {242601} (\bibinfo
  {year} {2016})}\BibitemShut {NoStop}%
\bibitem [{\citenamefont {Mills}\ \emph {et~al.}(2015)\citenamefont {Mills},
  \citenamefont {Shen}, \citenamefont {Xu},\ and\ \citenamefont
  {Liu}}]{Mills2015}%
  \BibitemOpen
  \bibfield  {author} {\bibinfo {author} {\bibfnamefont {S.~A.}\ \bibnamefont
  {Mills}}, \bibinfo {author} {\bibfnamefont {C.}~\bibnamefont {Shen}},
  \bibinfo {author} {\bibfnamefont {Z.}~\bibnamefont {Xu}}, \ and\ \bibinfo
  {author} {\bibfnamefont {Y.}~\bibnamefont {Liu}},\ }\href {\doibase
  10.1103/PhysRevB.92.144502} {\bibfield  {journal} {\bibinfo  {journal} {Phys.
  Rev. B}\ }\textbf {\bibinfo {volume} {92}},\ \bibinfo {pages} {144502}
  (\bibinfo {year} {2015})}\BibitemShut {NoStop}%
\bibitem [{\citenamefont {de~Souza~Silva}\ \emph {et~al.}(2006)\citenamefont
  {de~Souza~Silva}, \citenamefont {Van~de Vondel}, \citenamefont {Morelle},\
  and\ \citenamefont {Moshchalkov}}]{Silva2006}%
  \BibitemOpen
  \bibfield  {author} {\bibinfo {author} {\bibfnamefont {C.~C.}\ \bibnamefont
  {de~Souza~Silva}}, \bibinfo {author} {\bibfnamefont {J.}~\bibnamefont {Van~de
  Vondel}}, \bibinfo {author} {\bibfnamefont {M.}~\bibnamefont {Morelle}}, \
  and\ \bibinfo {author} {\bibfnamefont {V.~V.}\ \bibnamefont {Moshchalkov}},\
  }\href {\doibase 10.1038/nature04595} {\bibfield  {journal} {\bibinfo
  {journal} {Nature}\ }\textbf {\bibinfo {volume} {440}},\ \bibinfo {pages}
  {651} (\bibinfo {year} {2006})}\BibitemShut {NoStop}%
\bibitem [{\citenamefont {Veshchunov}\ \emph {et~al.}(2016)\citenamefont
  {Veshchunov}, \citenamefont {Magrini}, \citenamefont {Mironov}, \citenamefont
  {Godin}, \citenamefont {Trebbia}, \citenamefont {Buzdin}, \citenamefont
  {Tamarat},\ and\ \citenamefont {Lounis}}]{Veshchunov2016}%
  \BibitemOpen
  \bibfield  {author} {\bibinfo {author} {\bibfnamefont {I.~S.}\ \bibnamefont
  {Veshchunov}}, \bibinfo {author} {\bibfnamefont {W.}~\bibnamefont {Magrini}},
  \bibinfo {author} {\bibfnamefont {S.}~\bibnamefont {Mironov}}, \bibinfo
  {author} {\bibfnamefont {A.}~\bibnamefont {Godin}}, \bibinfo {author}
  {\bibfnamefont {J.-B.}\ \bibnamefont {Trebbia}}, \bibinfo {author}
  {\bibfnamefont {A.~I.}\ \bibnamefont {Buzdin}}, \bibinfo {author}
  {\bibfnamefont {P.}~\bibnamefont {Tamarat}}, \ and\ \bibinfo {author}
  {\bibfnamefont {B.}~\bibnamefont {Lounis}},\ }\href {\doibase
  10.1038/ncomms12801} {\bibfield  {journal} {\bibinfo  {journal} {Nature
  communications}\ }\textbf {\bibinfo {volume} {7}},\ \bibinfo {pages} {12801}
  (\bibinfo {year} {2016})}\BibitemShut {NoStop}%
\bibitem [{\citenamefont {Kremen}\ \emph {et~al.}(2016)\citenamefont {Kremen},
  \citenamefont {Wissberg}, \citenamefont {Haham}, \citenamefont {Persky},
  \citenamefont {Frenkel},\ and\ \citenamefont {Kalisky}}]{Kremen2016}%
  \BibitemOpen
  \bibfield  {author} {\bibinfo {author} {\bibfnamefont {A.}~\bibnamefont
  {Kremen}}, \bibinfo {author} {\bibfnamefont {S.}~\bibnamefont {Wissberg}},
  \bibinfo {author} {\bibfnamefont {N.}~\bibnamefont {Haham}}, \bibinfo
  {author} {\bibfnamefont {E.}~\bibnamefont {Persky}}, \bibinfo {author}
  {\bibfnamefont {Y.}~\bibnamefont {Frenkel}}, \ and\ \bibinfo {author}
  {\bibfnamefont {B.}~\bibnamefont {Kalisky}},\ }\href {\doibase
  10.1021/acs.nanolett.5b04444} {\bibfield  {journal} {\bibinfo  {journal}
  {Nano Letters}\ }\textbf {\bibinfo {volume} {16}},\ \bibinfo {pages} {1626}
  (\bibinfo {year} {2016})}\BibitemShut {NoStop}%
\bibitem [{\citenamefont {Gardner}\ \emph {et~al.}(2002)\citenamefont
  {Gardner}, \citenamefont {Wynn}, \citenamefont {Bonn}, \citenamefont {Liang},
  \citenamefont {Hardy}, \citenamefont {Kirtley}, \citenamefont {Kogan},\ and\
  \citenamefont {Moler}}]{Gardner2002}%
  \BibitemOpen
  \bibfield  {author} {\bibinfo {author} {\bibfnamefont {B.~W.}\ \bibnamefont
  {Gardner}}, \bibinfo {author} {\bibfnamefont {J.~C.}\ \bibnamefont {Wynn}},
  \bibinfo {author} {\bibfnamefont {D.~A.}\ \bibnamefont {Bonn}}, \bibinfo
  {author} {\bibfnamefont {R.}~\bibnamefont {Liang}}, \bibinfo {author}
  {\bibfnamefont {W.~N.}\ \bibnamefont {Hardy}}, \bibinfo {author}
  {\bibfnamefont {J.~R.}\ \bibnamefont {Kirtley}}, \bibinfo {author}
  {\bibfnamefont {V.~G.}\ \bibnamefont {Kogan}}, \ and\ \bibinfo {author}
  {\bibfnamefont {K.~A.}\ \bibnamefont {Moler}},\ }\href {\doibase
  http://dx.doi.org/10.1063/1.1445468} {\bibfield  {journal} {\bibinfo
  {journal} {Applied Physics Letters}\ }\textbf {\bibinfo {volume} {80}},\
  \bibinfo {pages} {1010} (\bibinfo {year} {2002})}\BibitemShut {NoStop}%
\bibitem [{\citenamefont {Straver}\ \emph {et~al.}(2008)\citenamefont
  {Straver}, \citenamefont {Hoffman}, \citenamefont {Auslaender}, \citenamefont
  {Rugar},\ and\ \citenamefont {Moler}}]{Straver2008}%
  \BibitemOpen
  \bibfield  {author} {\bibinfo {author} {\bibfnamefont {E.~W.~J.}\
  \bibnamefont {Straver}}, \bibinfo {author} {\bibfnamefont {J.~E.}\
  \bibnamefont {Hoffman}}, \bibinfo {author} {\bibfnamefont {O.~M.}\
  \bibnamefont {Auslaender}}, \bibinfo {author} {\bibfnamefont
  {D.}~\bibnamefont {Rugar}}, \ and\ \bibinfo {author} {\bibfnamefont {K.~A.}\
  \bibnamefont {Moler}},\ }\href {\doibase 10.1063/1.3000963} {\bibfield
  {journal} {\bibinfo  {journal} {Applied Physics Letters}\ }\textbf {\bibinfo
  {volume} {93}} (\bibinfo {year} {2008}),\ 10.1063/1.3000963}\BibitemShut
  {NoStop}%
\bibitem [{\citenamefont {Auslaender}\ \emph {et~al.}(2009)\citenamefont
  {Auslaender}, \citenamefont {Luan}, \citenamefont {Straver}, \citenamefont
  {Hoffman}, \citenamefont {Koshnick}, \citenamefont {Zeldov}, \citenamefont
  {Bonn}, \citenamefont {Liang}, \citenamefont {Hardy},\ and\ \citenamefont
  {Moler}}]{Auslaender2009}%
  \BibitemOpen
  \bibfield  {author} {\bibinfo {author} {\bibfnamefont {O.~M.}\ \bibnamefont
  {Auslaender}}, \bibinfo {author} {\bibfnamefont {L.}~\bibnamefont {Luan}},
  \bibinfo {author} {\bibfnamefont {E.~W.}\ \bibnamefont {Straver}}, \bibinfo
  {author} {\bibfnamefont {J.~E.}\ \bibnamefont {Hoffman}}, \bibinfo {author}
  {\bibfnamefont {N.~C.}\ \bibnamefont {Koshnick}}, \bibinfo {author}
  {\bibfnamefont {E.}~\bibnamefont {Zeldov}}, \bibinfo {author} {\bibfnamefont
  {D.~A.}\ \bibnamefont {Bonn}}, \bibinfo {author} {\bibfnamefont
  {R.}~\bibnamefont {Liang}}, \bibinfo {author} {\bibfnamefont {W.~N.}\
  \bibnamefont {Hardy}}, \ and\ \bibinfo {author} {\bibfnamefont {K.~A.}\
  \bibnamefont {Moler}},\ }\href {\doibase 10.1038/nphys1127} {\bibfield
  {journal} {\bibinfo  {journal} {Nature Physics}\ }\textbf {\bibinfo {volume}
  {5}},\ \bibinfo {pages} {35} (\bibinfo {year} {2009})}\BibitemShut {NoStop}%
\bibitem [{\citenamefont {Ge}\ \emph {et~al.}(2017)\citenamefont {Ge},
  \citenamefont {Gladilin}, \citenamefont {Tempere}, \citenamefont {Devreese},\
  and\ \citenamefont {Moshchalkov}}]{ge_controlled_2017}%
  \BibitemOpen
  \bibfield  {author} {\bibinfo {author} {\bibfnamefont {J.-Y.}\ \bibnamefont
  {Ge}}, \bibinfo {author} {\bibfnamefont {V.~N.}\ \bibnamefont {Gladilin}},
  \bibinfo {author} {\bibfnamefont {J.}~\bibnamefont {Tempere}}, \bibinfo
  {author} {\bibfnamefont {J.}~\bibnamefont {Devreese}}, \ and\ \bibinfo
  {author} {\bibfnamefont {V.~V.}\ \bibnamefont {Moshchalkov}},\ }\href
  {\doibase 10.1021/acs.nanolett.7b02180} {\bibfield  {journal} {\bibinfo
  {journal} {Nano Letters}\ }\textbf {\bibinfo {volume} {17}},\ \bibinfo
  {pages} {5003} (\bibinfo {year} {2017})}\BibitemShut {NoStop}%
\bibitem [{\citenamefont {Ge}\ \emph {et~al.}(2016)\citenamefont {Ge},
  \citenamefont {Gladilin}, \citenamefont {Tempere}, \citenamefont {Xue},
  \citenamefont {Devreese}, \citenamefont {Van~de Vondel}, \citenamefont
  {Zhou},\ and\ \citenamefont {Moshchalkov}}]{ge2016nanoscale}%
  \BibitemOpen
  \bibfield  {author} {\bibinfo {author} {\bibfnamefont {J.-Y.}\ \bibnamefont
  {Ge}}, \bibinfo {author} {\bibfnamefont {V.~N.}\ \bibnamefont {Gladilin}},
  \bibinfo {author} {\bibfnamefont {J.}~\bibnamefont {Tempere}}, \bibinfo
  {author} {\bibfnamefont {C.}~\bibnamefont {Xue}}, \bibinfo {author}
  {\bibfnamefont {J.~T.}\ \bibnamefont {Devreese}}, \bibinfo {author}
  {\bibfnamefont {J.}~\bibnamefont {Van~de Vondel}}, \bibinfo {author}
  {\bibfnamefont {Y.}~\bibnamefont {Zhou}}, \ and\ \bibinfo {author}
  {\bibfnamefont {V.~V.}\ \bibnamefont {Moshchalkov}},\ }\href {\doibase
  10.1038/ncomms13880} {\bibfield  {journal} {\bibinfo  {journal} {Nature
  communications}\ }\textbf {\bibinfo {volume} {7}},\ \bibinfo {pages} {13880}
  (\bibinfo {year} {2016})}\BibitemShut {NoStop}%
\bibitem [{\citenamefont {Ma}\ \emph {et~al.}(2018)\citenamefont {Ma},
  \citenamefont {Reichhardt},\ and\ \citenamefont
  {Reichhardt}}]{xiaoyu_manipulation_2018}%
  \BibitemOpen
  \bibfield  {author} {\bibinfo {author} {\bibfnamefont {X.}~\bibnamefont
  {Ma}}, \bibinfo {author} {\bibfnamefont {C.~J.~O.}\ \bibnamefont
  {Reichhardt}}, \ and\ \bibinfo {author} {\bibfnamefont {C.}~\bibnamefont
  {Reichhardt}},\ }\href {\doibase 10.1103/PhysRevB.97.214521} {\bibfield
  {journal} {\bibinfo  {journal} {Phys. Rev. B}\ }\textbf {\bibinfo {volume}
  {97}},\ \bibinfo {pages} {214521} (\bibinfo {year} {2018})}\BibitemShut
  {NoStop}%
\bibitem [{\citenamefont {Milo{\v{s}}evi{\'c}}\ and\ \citenamefont
  {Peeters}(2010)}]{Milosevic2010}%
  \BibitemOpen
  \bibfield  {author} {\bibinfo {author} {\bibfnamefont {M.}~\bibnamefont
  {Milo{\v{s}}evi{\'c}}}\ and\ \bibinfo {author} {\bibfnamefont
  {F.}~\bibnamefont {Peeters}},\ }\href {\doibase 10.1063/1.3425672} {\bibfield
   {journal} {\bibinfo  {journal} {Applied Physics Letters}\ }\textbf {\bibinfo
  {volume} {96}},\ \bibinfo {pages} {192501} (\bibinfo {year}
  {2010})}\BibitemShut {NoStop}%
\bibitem [{\citenamefont {Polshyn}\ \emph {et~al.}(2018)\citenamefont
  {Polshyn}, \citenamefont {Naibert},\ and\ \citenamefont
  {Budakian}}]{Polshyn_2017}%
  \BibitemOpen
  \bibfield  {author} {\bibinfo {author} {\bibfnamefont {H.}~\bibnamefont
  {Polshyn}}, \bibinfo {author} {\bibfnamefont {T.~R.}\ \bibnamefont
  {Naibert}}, \ and\ \bibinfo {author} {\bibfnamefont {R.}~\bibnamefont
  {Budakian}},\ }\href {\doibase 10.1103/PhysRevB.97.184501} {\bibfield
  {journal} {\bibinfo  {journal} {Phys. Rev. B}\ }\textbf {\bibinfo {volume}
  {97}},\ \bibinfo {pages} {184501} (\bibinfo {year} {2018})}\BibitemShut
  {NoStop}%
\bibitem [{\citenamefont {Polshyn}(2017)}]{polshyn_magnetic_2017}%
  \BibitemOpen
  \bibfield  {author} {\bibinfo {author} {\bibfnamefont {H.}~\bibnamefont
  {Polshyn}},\ }\emph {\bibinfo {title} {Magnetic force microscopy studies of
  mesoscopic superconducting structures}},\ \href@noop {} {Ph.D. thesis},\
  \bibinfo  {school} {University of Illinois at Urbana-Champaign} (\bibinfo
  {year} {2017})\BibitemShut {NoStop}%
\bibitem [{SI()}]{SI}%
  \BibitemOpen
  \href@noop {} {\ }\bibinfo {note} {See the supporting
  information.}\BibitemShut {Stop}%
\bibitem [{\citenamefont {Albrecht}\ \emph {et~al.}(1991)\citenamefont
  {Albrecht}, \citenamefont {Grütter}, \citenamefont {Horne},\ and\
  \citenamefont {Rugar}}]{Albrecht1991}%
  \BibitemOpen
  \bibfield  {author} {\bibinfo {author} {\bibfnamefont {T.~R.}\ \bibnamefont
  {Albrecht}}, \bibinfo {author} {\bibfnamefont {P.}~\bibnamefont {Grütter}},
  \bibinfo {author} {\bibfnamefont {D.}~\bibnamefont {Horne}}, \ and\ \bibinfo
  {author} {\bibfnamefont {D.}~\bibnamefont {Rugar}},\ }\href {\doibase
  http://dx.doi.org/10.1063/1.347347} {\bibfield  {journal} {\bibinfo
  {journal} {Journal of Applied Physics}\ }\textbf {\bibinfo {volume} {69}},\
  \bibinfo {pages} {668} (\bibinfo {year} {1991})}\BibitemShut {NoStop}%
\bibitem [{\citenamefont {van~der Wiel}\ \emph {et~al.}(2002)\citenamefont
  {van~der Wiel}, \citenamefont {De~Franceschi}, \citenamefont {Elzerman},
  \citenamefont {Fujisawa}, \citenamefont {Tarucha},\ and\ \citenamefont
  {Kouwenhoven}}]{Wiel2002}%
  \BibitemOpen
  \bibfield  {author} {\bibinfo {author} {\bibfnamefont {W.~G.}\ \bibnamefont
  {van~der Wiel}}, \bibinfo {author} {\bibfnamefont {S.}~\bibnamefont
  {De~Franceschi}}, \bibinfo {author} {\bibfnamefont {J.~M.}\ \bibnamefont
  {Elzerman}}, \bibinfo {author} {\bibfnamefont {T.}~\bibnamefont {Fujisawa}},
  \bibinfo {author} {\bibfnamefont {S.}~\bibnamefont {Tarucha}}, \ and\
  \bibinfo {author} {\bibfnamefont {L.~P.}\ \bibnamefont {Kouwenhoven}},\
  }\href {\doibase 10.1103/RevModPhys.75.1} {\bibfield  {journal} {\bibinfo
  {journal} {Rev. Mod. Phys.}\ }\textbf {\bibinfo {volume} {75}},\ \bibinfo
  {pages} {1} (\bibinfo {year} {2002})}\BibitemShut {NoStop}%
\bibitem [{\citenamefont {Little}(1967)}]{Little1967}%
  \BibitemOpen
  \bibfield  {author} {\bibinfo {author} {\bibfnamefont {W.~A.}\ \bibnamefont
  {Little}},\ }\href {\doibase 10.1103/PhysRev.156.396} {\bibfield  {journal}
  {\bibinfo  {journal} {Phys. Rev.}\ }\textbf {\bibinfo {volume} {156}},\
  \bibinfo {pages} {396} (\bibinfo {year} {1967})}\BibitemShut {NoStop}%
\bibitem [{\citenamefont {Langer}\ and\ \citenamefont
  {Ambegaokar}(1967)}]{Langer1967}%
  \BibitemOpen
  \bibfield  {author} {\bibinfo {author} {\bibfnamefont {J.~S.}\ \bibnamefont
  {Langer}}\ and\ \bibinfo {author} {\bibfnamefont {V.}~\bibnamefont
  {Ambegaokar}},\ }\href {\doibase 10.1103/PhysRev.164.498} {\bibfield
  {journal} {\bibinfo  {journal} {Phys. Rev.}\ }\textbf {\bibinfo {volume}
  {164}},\ \bibinfo {pages} {498} (\bibinfo {year} {1967})}\BibitemShut
  {NoStop}%
\end{thebibliography}
%

\clearpage
\pagebreak
\widetext

\setcounter{equation}{0}
\setcounter{figure}{0}
\setcounter{table}{0}
\makeatletter
\renewcommand{\theequation}{S\arabic{equation}}
\renewcommand{\thefigure}{S\arabic{figure}}
\renewcommand{\bibnumfmt}[1]{[S#1]}

\section{Supporting information}

\subsection{Experimental procedures}

\begin{figure}[!h]
\includegraphics[]{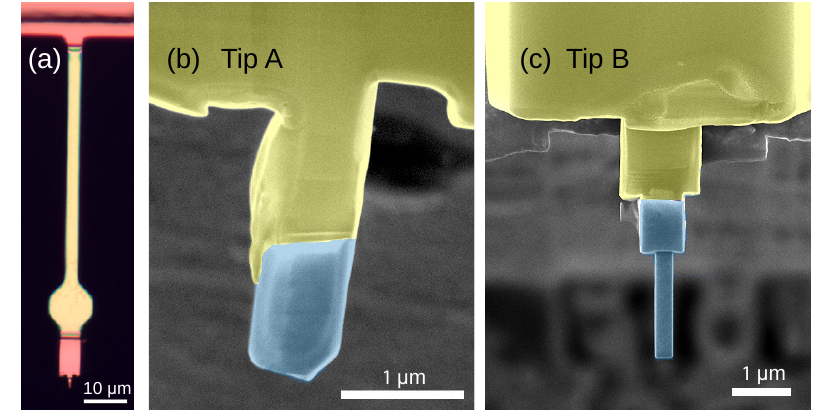}
\caption[] {(a) Si cantilever (b) Tip~A (c) Tip~B
}
\label{FigS1}
\end{figure}

In our measurements we use custom made ultra-soft silicon cantilevers with $\mathrm{Sm Co_5}$ particles mounted at the tip, shown in Fig.~\ref{FigS1}.
To fabricate them, we first attach a $\mathrm{Sm Co_5}$ magnetic particle of appropriate size to the tip of the cantilever using a micromanipulator.
G1 epoxy (Gatan Inc.) is used to glue the particle.
The magnetic moment of the particle is aligned with the axis of the cantilever by applying a magnetic field while epoxy is being cured.
Next, the particle is trimmed to a desired regular shape using a focused ion beam (FIB). To avoid the ion damage of the magnetic particle, we use low ion currents: 40-20 pA for the rough cuts and 1 pA for the finishing cuts. 
The magnetic moment of the particle is characterized using cantilever torque magnetometry.

For measurements reported here we use two MFM tips (Fig.~\ref{FigS1}). Tip~A is used to measure Ring~1,3 and 4, while Tip~B is used to measure Ring~2. 
Tip A has cantilever length $L=110~\mathrm{\mu m}$, resonant frequency $f_0\simeq 4146$~Hz, spring constant $k=1.1 \times  10^{-4}$~N/m and quality factor $Q\simeq 10^3$.  The $\mathrm{Sm Co_5}$ particle of tip A has magnetic moment $m_{\mathrm{tip},\parallel}=2.2 \cdot 10^{-13}$ J/T. 
Tip B is mounted on a cantilever with length $L=80~\mathrm{\mu m}$, resonant frequency $f_0\simeq 7675$~Hz, spring constant $k=1.8 \times  10^{-4}$~N/m and quality factor $Q\simeq 31,800$ at 4~K. The $\mathrm{Sm Co_5}$ particle has  magnetic moment with components $m_{\mathrm{tip},\parallel}=7.2 \cdot 10^{-13}$ J/T and $m_{\mathrm{tip},\perp}=3 \cdot 10^{-14}$ J/T.

\begin{figure}[t]
\includegraphics[]{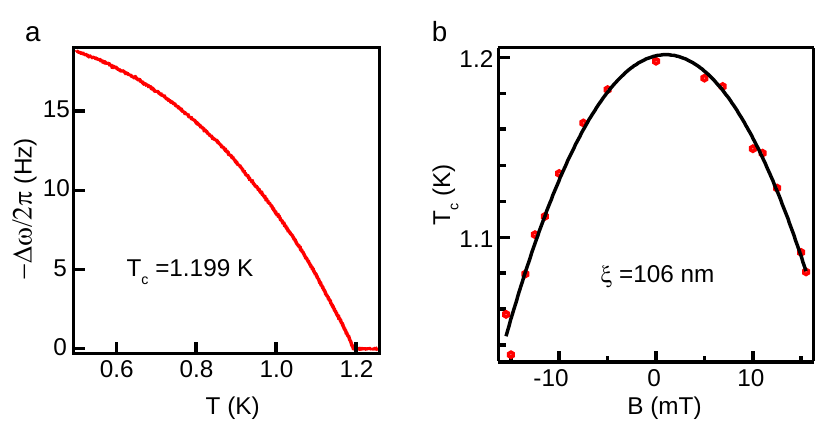}
\caption[] {(a) Resonance frequency shift measured  above Ring~2 as a function of temperature shows clear onset of superconductivity at  1.199~K. (b) Suppression of the superconducting transition by out-of-plane magnetic field in Ring~2 used to determine the coherence length $\xi(0)$.
}
\label{SCparameters}
\end{figure}

The superconducting parameters of the measured  structures are measured using the same MFM setup. The superconducting transition temperature $T_c$ is measured by  parking a MFM tip above the structure and monitoring the resonance frequency shift as a function of temperature (Fig.~\ref{SCparameters}~a).
The superconducting coherence length $\xi (0)$  is determined from the suppression of the superconducting transition by external magnetic field (Fig.~\ref{SCparameters})b.
The critical field of a thin wall ring perpendicular to the plane of the ring is given by~\cite{Polshyn_2017}
\[
B_c(T)=B_c(0)\sqrt{1-T/T_c},
\]
with
\[B_c(0)=\frac{\sqrt{3}}{\pi}\frac{ \Phi_0}{w \xi(0)},\]
where $w$ is the width of the wall of the ring. 
We obtain $\xi(0)$ from the known value of $w$ and the value of $B_c(0)$ determined from the fit (see Fig.~\ref{SCparameters}b).
For Ring 2 with wall width  $w$=236~nm we find $\xi(0)=106$~nm.
From the temperature dependence $\xi(T)= \xi(0)/\sqrt{1-T/T_c}$, we estimate that at temperatures  used in this work ($T/T_c>0.97$) the superconducting coherence length $\xi>630$~nm, which is much larger than the width and thickness of the wall of the studied structures. The same is also true for superconducting penetration depth, which we estimate to be $\lambda >1$~$\mathrm{\mu m}$ using $\lambda(T)=(1-T/T_c)^{-0.5}\times 166$~nm obtained for similar aluminium structures~\cite{Polshyn_2017}. Thus, the structures are  in the limit of 1D superconducting networks with negligible magnetic screening.
The parameters of the superconducting structures used in experiments are shown in Tab.~\ref{Tab1}.

\begin {table}
\caption{\label{Tab1}Parameters of structures used in experiments.}
\begin{tabular}{|c|c|c|c|c|}
\hline
Ring & R($\mathrm{\mu}$m)  & w(nm) & $T_c$(K)\\
\hline
1& 0.95 & 133 & $\sim 0.95$ \\
2& 1.99 & 236 & 1.199 \\
3& 0.94 & 130 &$\sim 0.98$ \\
4& 0.94 & 125 & $\sim 0.98$ \\
\hline
\end{tabular}
\end {table}

\subsection{Energetics of vortex states in multi-loop superconducting structures}
Here we derive the energy of multi-loop superconducting structures
in which the vortices can sit only inside the loops. 
A vortex state of such structure, comprised of $N$ elementary loops, can be described by $N$ winding numbers $\hat{n}=(n_1,\dots,n_N)$ -- the numbers of vortices hosted by each loop.
We assume, 
that the walls of the wires, comprising the structures, are narrow and thin ($w,t\ll\lambda, \xi$) and the direct effect of the magnetic field of the MFM tip on the wires can be neglected.
In this case, the free energy of the vortex states, hosted by the structure, depends only on the fluxes threading the	 elementary loops $\hat{\phi}=(\phi_1, \dots,\phi_N)$, rather than on the full spatial distribution of the magnetic field:
\begin{equation}
F=F(\hat{\phi},\hat{n}).
\label{eq:}
\end{equation}

Let us consider a superconducting wire network with  $N_{\mathrm{loops}}$ elementary loops, $N_{\mathrm{links}}$ wires and $N_{\mathrm{nodes}}$ nodes.
It can be shown that  $N_{\mathrm{links}}= N_{\mathrm{loops}}+N_{\mathrm{nodes}}-1$.
The gauge-invariant phase difference for each wire is given by:
 \begin{equation}\label{eq:GIPhase}
\gamma_k \equiv \Delta \varphi_k -(2 \pi /\Phi_0) \int_{\text{k-th wire}} \mathbf{A}\cdot d\mathbf{s}. 
\end{equation}
Two types of the equations, similar to Kirchoff's rules, can be written for the network. 
The first  type follows from the fluxoid quantization condition\cite{Tinkham1996} applied to each elementary $i$-th loop: \begin{equation}\label{eq:Phasewinding}
\sum_{j:\,i\text{-th loop}}{\gamma_j}=2\pi(n_i-\phi_i).
\end{equation}
The second type of the equations is obtained from the conservation of current in each node, for  $p$-th node:
\begin{equation}\label{eq:CurrentConservation}
\sum_{j:\,p\text{-th node}}{I_j}=0,
\end{equation}
where  $I_j$ is the current in the $j$-th wire. 
The sum in Eq.~\eqref{eq:CurrentConservation} is taken over the wires connected to the $p$-th node, respecting the chosen orientation of the wires in the network.
There are $N_{\mathrm{nodes}}-1$ independent equations of the second type.

If the wires forming the network are long in comparison to the superconducting coherence length $\xi$, the pairbreaking effects are small, thus further we assume homogeneous  superfluid density described by penetration depth $\lambda$.
Under this condition, the current $I_j$ can be expressed through $\gamma_j$  using London equation as:
\begin{equation}\label{eq:CurrentDensityWire}
I_j=\frac{1}{\mu_0 \lambda^2} \left( \frac{\Phi_0}{2\pi}\right) \frac{\gamma_k}{l_j} S_j,
\end{equation}
where $l_j$ and $S_j$ are the length and the cross sectional area of the wire $j$.
It is convenient to introduce the kinetic inductance $L_j^K$ of a wire with length $l_j$ and cross-sectional area $S_j$ as follows:
\begin{equation}\label{eq:KineticInductance}
L_j^K=\mu_0 \lambda^2 \frac {l_j}{S_j}.
\end{equation}
Combining equations \eqref{eq:CurrentDensityWire} and \eqref{eq:KineticInductance} yields
\begin{equation}\label{eq:WireCurrent}
I_j=\frac{\Phi_0}{2 \pi} \frac{1}{L^K_j}\, \gamma_j.
\end{equation}
Replacing $I_j$'s in equation\eqref{eq:CurrentConservation} we obtain
\begin{equation}\label{eq:CurrentConserv2}
\sum_{j:\,p\text{-th node}}\,{\frac{\gamma_j}{L_j^K}}=0.
\end{equation}
Eq.~\eqref{eq:CurrentConserv2}, written for $(N_{\mathrm{nodes}}-1)$ nodes of the network, together with  Eq.~\eqref{eq:Phasewinding}, written for $N_{\mathrm{loops}}$ loops form a system of $N_{\mathrm{links}}$ linear  equations.
Solving this system allows us to find the gauge-invariant phase differences $\gamma_k$ across all the wires of the network.
From equations \eqref{eq:CurrentConserv2} and \eqref{eq:Phasewinding}, it is easy to see, that $\gamma$'s depend linearly on $(\hat{n}-\hat{\phi})$.

\begin{figure}[b]
\centering
\includegraphics[]{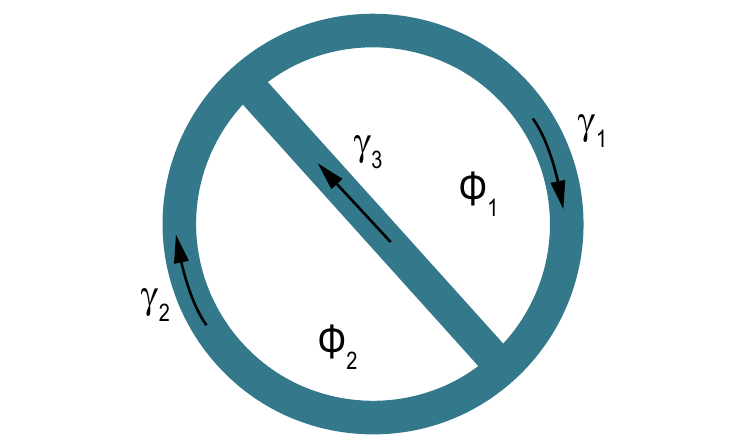}
\caption []{Orientation of links in ring with crossbar.}
\label{fig:crossbarRingDiagram}
\end{figure}

For a wire  of small cross-section ($S\ll \lambda^2$), the energy $E^M$ due to the magnetic field  is negligible in comparison to the kinetic energy of the supercurrent $E^K$.
Thus, the energy of  the $j$-th wire is   
\begin{equation}\label{eq:wireEnergy}
E_j\approx E_j^K=\frac {L_j^K I_j^2}{2 }=\left(\frac{\Phi_0}{2\pi}\right)^2\frac {\gamma_j^2}{2 L_j^K}.
\end{equation}
The total energy of the superconducting network is obtained by summing the energies of all links:
\begin{equation}
E=\sum_{j}^{N_{\mathrm{links}}}{E_j(\gamma_j)}=\left(\frac{\Phi_0}{2\pi}\right)^2\,
\sum_{j}^{N_{\mathrm{links}}}{\frac {\gamma_j^2}{2 L^K}}.
\label{eq:TotalEnergy}
\end{equation}
Since the $\gamma$'s depend linearly on $(\hat{n}-\hat{\phi})$, the total energy of the network is a quadratic form of $(\hat{n}-\hat{\phi})$.   
\begin{figure*}[!h]
\centering
\includegraphics[]{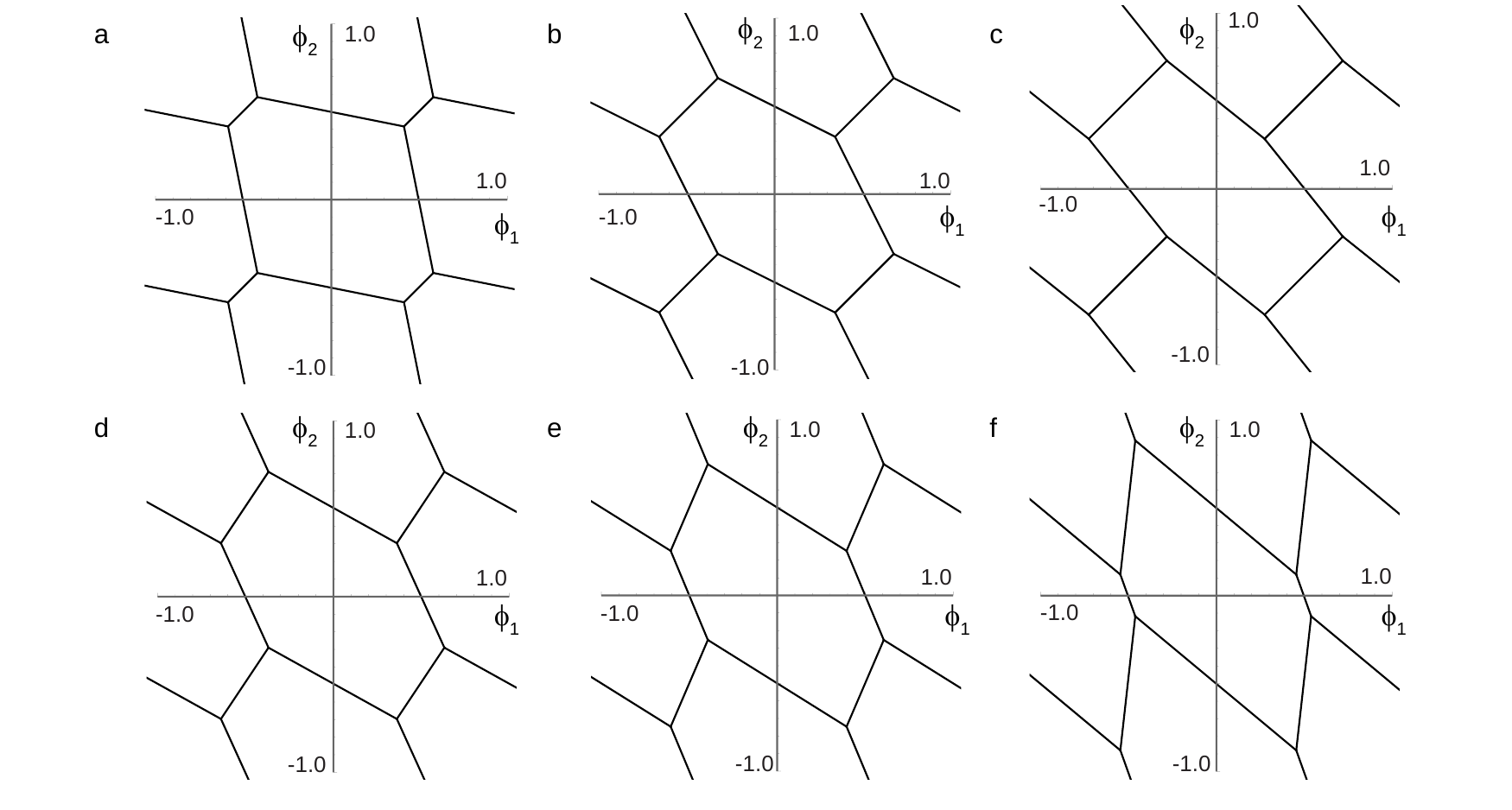}
\caption []{Stability diagrams of the vortex states in two-loop structures with different  $L^K_1$, $L^K_2$ and $L^K_3$. The diagrams, shown in (a-c), correspond to symmetric structures $L^K_1=L^K_2$, while the diagrams shown in (d-e) correspond to $L^K_1\ne L^K_2$. (a) Weak inter-loop coupling: $L^K_1 :L^K_2 : L^K_3=1:1:0.25$; (b) moderate inter-loop coupling: $L^K_1 :L^K_2 : L^K_3=1:1:1$; (c) strong inter-loop coupling:$L^K_1 :L^K_2 : L^K_3=1:1:4$; (d) $L^K_1 :L^K_2 : L^K_3=0.8:1.2:1$;  (e) $L^K_1 :L^K_2 : L^K_3=0.6:1.4:1$; (f) $L^K_1 :L^K_2 : L^K_3=0.2:1.8:1$.}
\label{HonecombDiagramsTheoryVariety}
\end{figure*}

A vortex state in a ring with a crossbar (Fig.~\ref{fig:crossbarRingDiagram}) is described by a pair of winding numbers $\{ n_1, n_2 \}$ for  the top and the bottom halves of the rings.
Writing down equations \ref{eq:Phasewinding}, \ref{eq:CurrentConserv2} for this structure (using the orientation of wires shown in Fig.~\ref{fig:crossbarRingDiagram}) yields a system of three equations:
\begin {equation}\label{eq:sytemCrossbar1}
\left\{
\begin{aligned}
&\gamma_1+\gamma_3=-2\pi \tilde{\phi_1}\\
&\gamma_2-\gamma_3=-2\pi \tilde{\phi_2} \\
&\frac{\gamma_1}{L_1^K}=\frac{\gamma_3}{L_3^K}+\frac{\gamma_2}{L_2^K}\\
\end{aligned}\right.
\end{equation}
where $\tilde{\phi_i}=\phi_i-n_i$.
The total energy of the structure is obtained from Eq.~\eqref{eq:TotalEnergy}:
\begin{equation}
E=\frac{1}{2} \Phi_0^2 L^K_{\Vert}\left[\frac{\tilde{\phi_1}^2}{L^K_1 L^K_3}+\frac{\tilde{\phi_2}^2}{L^K_2 L^K_3}+\frac{(\tilde{\phi_1} +\tilde{\phi_2})^2}{L^K_1 L^K_2}\right],
\label{eq:GeneralEnergy}
\end{equation}
where, for convenience, we introduce $
1/L_{\Vert}^{K}=1/L^{K}_1+1/L^{K}_2+1/L_3^{K}$.
In the case of a symmetric structure, such that $L_1^K=L_2^K=L^K $, the energy of the structure is
\begin{equation}
E=\frac{1}{2} \frac {\Phi_0^2} {L^K (\beta +1)}\left[\tilde{\phi_1}^2+2\beta \tilde{\phi_1}\tilde{\phi_2}+\tilde{\phi_2}^2\right],
\label{eq:SymmetricRing}
\end{equation}
where $\beta=L^K_3/(L^K_1+L^K_3)$ is a parameter that characterises the strength of coupling of the two halves. For a ring with a crossbar made of wires with a homogeneous cross section, shown in Fig.~\ref{fig:crossbarRingDiagram}, $\beta=2/(2+\pi)\approx0.39$.
It is instructive to consider two special limits: if $\beta = 0$, equation~\eqref{eq:SymmetricRing} yields $E\propto \tilde{\phi_1}^2+\tilde{\phi_2}^2$ -- the structure behaves like two isolated rings; if $\beta =1$: $E\propto (\tilde{\phi_1}+\tilde{\phi_2})^2$, which corresponds to a "vanishing" crossbar, and the structures behaves like a single ring.

The stability region of the vortex state with vortex numbers $\{n_1,n_2\}$  has a hexagonal shape in ($\phi_1, \phi_2$) coordinates, set  by the following constrains:
\begin{equation}
\left\{
\begin{aligned}
&\left|\tilde{\phi_1}+\beta \tilde{\phi_2}\right|<  1/2\\
&\left|\beta\tilde{\phi_1}+ \tilde{\phi_2}\right|<  1/2\\
&|\tilde{\phi_1}-\tilde{\phi_2}|< 1\\
\end{aligned}
\right.
\label{eq:RegionOfStability}
\end{equation}
The examples of honeycomb stability diagram for structures with  different values of inter-loop coupling are shown in Fig.~\ref{HonecombDiagramsTheoryVariety}(a-c).
Increasing the kinetic inductance of the crossbar increases the inter-loop coupling, which results in a squeezing  of the  hexagons in  (1,1) direction.
The depth of the modulation of the honeycomb cells $\beta$, measured as shown in Fig.~2, can be expressed in terms of  the inter-loop coupling constant $\beta$: 
\begin{equation}
\beta= \frac{l_2}{l_1}
\label{eq:HonecombModulation}
\end{equation}
Thus, the coupling $\beta$ between two halves of the structure can  be determined  from the  observed pattern of vortex transitions.
In asymmetric structures $(L^K_1 \ne L^K_2)$, the hexagons in the stability diagram become oblique, as shown in Fig.~\ref{HonecombDiagramsTheoryVariety}(d-f).

\subsection{Matching of the vortex transitions by simulating magnetic field of the tip}

In order to quantitatively understand  the pattern of vortex transition observed in Fig.~2 we  simulate them using a model for magnetic tip. 
As a first approximation, we use the dimensions of the magnetic particle measured with SEM (600 x 600 x 1000~nm$^3$) and assume the homogeneous distribution of the magnetic moment $m_{\Vert} =2.2\times10^{-13}$~J/T, that is measured using cantilever magnetometry.
Further, we calculate the fluxes $(\phi_1, \phi_2)$ as a function of the tip position $(x_{tip}, y_{tip},h)$ and use Eqs.~\eqref{eq:RegionOfStability} to  find the positions of vortex transitions. Next, we use $m_{\Vert}$ and $h$ as tuning parameters of the model to match the observed transitions. A good match is obtained  with $m_{\Vert}=2.43\times10^{-13}$~J/T and  $h=1.1$~$\mathrm{\mu}m$. Finally, we use the refined model of the tip to calculate the field distribution $B_z(x,y)$ and the modulation $\delta B_z(x,y)= (\partial B_z/\partial y)\cdot \delta y$ due to tip oscillation along $y$-axis with amplitude $x_0$=5~nm (see Figs.2 f-g). 

\subsection{Stochastic resonance imaging of the transition rates}

Here we generalize the model of the interaction of MFM cantilever with  thermally-activated transitions between vortex states in a ring (Ref.~\cite{Polshyn_2017}) to a more general case of a small superconducting network composed of narrow superconducting wires.
We consider two vortex configurations \textit{a} and \textit{b}  near the transition point at $\hat{\phi_0}$, where $F_a(\hat{\phi_0})=F_b(\hat{\phi_0})$, and assume that all other vortex states  have much lower or higher energies and hence do not contribute to the  dynamics.
We assume that the flux modulation $\delta \hat{\phi}$ is sufficiently small so that 
$|F_{a}(\hat{\phi})-F_b(\hat{\phi}))|\lesssim k_B T$ and hence both vortex states have a substantial probability of being occupied. 
We also assume that the energy barrier for transition between states $a$ and $b$ is sufficiently low to permit thermally activated transitions.

In the regime described above, the state of the structure exhibit thermally driven fluctuations between the two lowest-energy vortex states of the ring ($a$ and $b$).
Consequently, the supercurrents $\hat{I}(t)$ circulating in the structure also contain a two-level stochastic component.
The dynamics of the system is governed by $\hat{\phi}$-dependent transition rates  $\Gamma_{a}$ and $\Gamma_{b}$  that correspond to transitions $a \to b$ and $b \to a$.
The probability to find the structure in state $a$, when it is in thermal equilibrium, and the cantilever is stationary, is given by
\begin{equation}
P^{eq}_a(\hat{\phi})=\Gamma_b/(\Gamma_a+\Gamma_b).
\end{equation}
The dynamics  of the probability $P_a(t)$ is determined by the relaxation rate $\nu_r=\Gamma_a+\Gamma_b$
\begin{align} 
dP_a/dt&=- \nu_r  P_a +\Gamma_{b}. \label{eq:Pdynamics}
\end{align}
 At $\hat{\phi}=\hat{\phi_0}$:  $\Gamma_{a}=\Gamma_{b}$, so that $P^{eq}_a(\hat{\phi_0})=0.5$ and $\nu_r=2 \Gamma_{a}$. 

The supercurrent in the structure exerts a force on the magnetic particle given by ${\zeta (t)=\hat{\kappa}(\mathbf{r}_{tip})\cdot \hat{I(t)}}$, where $\hat{\kappa}(\mathbf{r}_{tip})$ represents the coupling between the circulating supercurrent and the cantilever. The $j$-th component of $\hat{\kappa}(\mathbf{r}_{tip})$ represents the coupling to the current $I_j$ flowing in the $j$-th wire of the structure.
The equation of motion for the cantilever becomes:
\begin{equation}\label{eq:CantileverDynamics}
\ddot x +2\gamma_0 \dot x+\omega_0^2 x = \frac{\omega_0^2}{k} [f(t) +\zeta (t,x)],
\end{equation}
where $x$ is the displacement of the tip from its equilibrium position, $\gamma_0$  is the unmodified dissipation of the cantilever, and $f(t)$ is the force applied by the feedback controller, which resonantly drives the cantilever at a fixed amplitude $x_0$.

Tip oscillations with amplitude $x_0$ generate a small modulation of the flux $\delta \hat{\phi}=(\partial\hat{\phi}/\partial x)\, x_0$, which  in turn modulates the transition rates $\Gamma_{a}$ and $ \Gamma_{b}$. The modulation of the transition  rates leads to statistical correlation between  $\zeta(t)$ and $x(t)$, which causes the shift of the resonant frequency and dissipation of the cantilever.
The  motion of cantilever can be represented as a sum of the coherent and stochastic terms: $x(t)=x_0 e^{ i\omega t}+x_s(t)$. The stochastic part of the motion has a vanishing time time-averaged Fourier component $\langle\hat{x_s} (\omega)\rangle=0$. 
Since we are mainly interested in the effect of the fluctuating force on the frequency and dissipation of the cantilever $\Delta \omega\equiv\left<\omega-\omega_0\right>$ and $\Delta \gamma\equiv \left<\gamma-\gamma_0\right>$, emerging due to correlations between  $\zeta(t)$ and $x(t)$, we consider only the case of weak stochastic force, i.e., $|\zeta (t,x)|\ll k x_0$. In this case $|x_s(t)|\ll x_0$, and we can approximate $x(t)\simeq x_0 e^{ i\omega t}$ in calculating $\langle\hat{ \zeta}(\omega)\rangle$. The crucial approximation of weak $|\zeta (t,x)|$, which enables us to effectively decouple the cantilever dynamics from the dynamics of the phase slips, is justified in our measurements, because the observed resonance frequency shifts are small.

For sufficiently weak flux modulation such that $(\partial P^{eq}_a/\partial\hat{\phi}) \delta \hat{\phi}   \ll 1$, the resulting modulation of $P_a(t)$ is linear in $\delta \hat{\phi}$, with
\begin{align}\label{eq:Psolution}
&P_a(t)\simeq P_a^{eq}(\hat{\phi_0}) +\delta P e^{i(\omega t-\theta)},\\
&\delta P=\frac{\partial P^{eq}_a}{\partial\hat{\phi}} \delta \hat{\phi}\cos{\theta}, \label{eq:PsolutionAmp}\\
&\theta=\arctan\left(\frac{\omega}{\nu_r}\right).\label{eq:PsolutionPhase}
\end{align}

The ensemble-averaged value of current is given by
\begin{equation}
\langle{ \hat{I}(t)}\rangle=\hat{I_a}(\hat{\phi}) P_a(t)+\hat{I_{b}}(\hat{\phi}) (1- P_a(t)).
\end{equation}
Expanding $\hat{I}$ around $\hat{\phi_0}$, $\hat{I}(\hat{\phi})=\hat{I}(\hat{\phi_0})+(\partial\hat{I}/\partial \hat{\phi}) \cdot \delta \hat{\phi}$  we obtain
\begin{equation}
\langle{ \hat{I}(t)}\rangle=[\hat{I_a}(\hat{\phi_0})-\hat{I_{b}}(\hat{\phi_0})]\delta P e^{i(\omega t-\theta)}+\frac 1 2 \frac{\partial(\hat{I_a}+\hat{I_b})}{\partial \hat{\phi}} \cdot \delta \hat{\phi}.
\label{eq:CurrentAvg}
\end{equation}
The first term in $\langle I(t) \rangle$ describes the contribution to the current from the thermally-activated transitions between the two states. The second term is  not relevant to the effect of interest  since it describes the flux dependence of the currents in each state. 
From Eq.~\ref{eq:CurrentAvg}, we find the Fourier component of the statistically-synchronized stochastic force due to the cantilever-driven phase slips  $\langle\hat{ \zeta}(\omega)\rangle=-\hat{\kappa}(\mathbf{r}_{tip}) \Delta \hat{I} \delta P e^{-i \theta}$, where $\Delta \hat{I}=\hat{I_a}(\hat{\phi_0})-\hat{I_{b}}(\hat{\phi_0})$.
Finally, $\langle\hat{ \zeta}(\omega)\rangle$ enables us to find $\Delta \omega$ and $\Delta \gamma$:
 \begin{align}
&\Delta \omega \simeq \frac{\omega_0}{2}\,  \frac{ \nu_r^2}{\nu_r^2+\omega_0^2} K, 
\label{eq:Frequency}\\
&\Delta \gamma = -\frac{\omega_0}{2}\,\frac{\omega_0\, \nu_r}{\nu_r^2+\omega_0^2}K, 
\label{eq:Damping}
\end{align}
where $K$ is a coupling constant given by
\begin{equation}
K=\left(\frac{\hat{\kappa}(\mathbf{r}_{tip})  }{k} \frac{d \hat{\phi}}{dx}\right)\left(\Delta \hat{I} (\hat{\phi}) \frac{\partial P_a^{eq}}{\partial\hat{\phi}}\right).
\label{eq:K}
\end{equation}
First factor in Eq.~\ref{eq:K}  represents the geometric part of the coupling and depends on the mutual position of the structure and the cantilever, while the second factor describes the dependence of $K$ on the strength of supercurrents in the structure.

\begin{figure*}[] 
\centering
\includegraphics[]{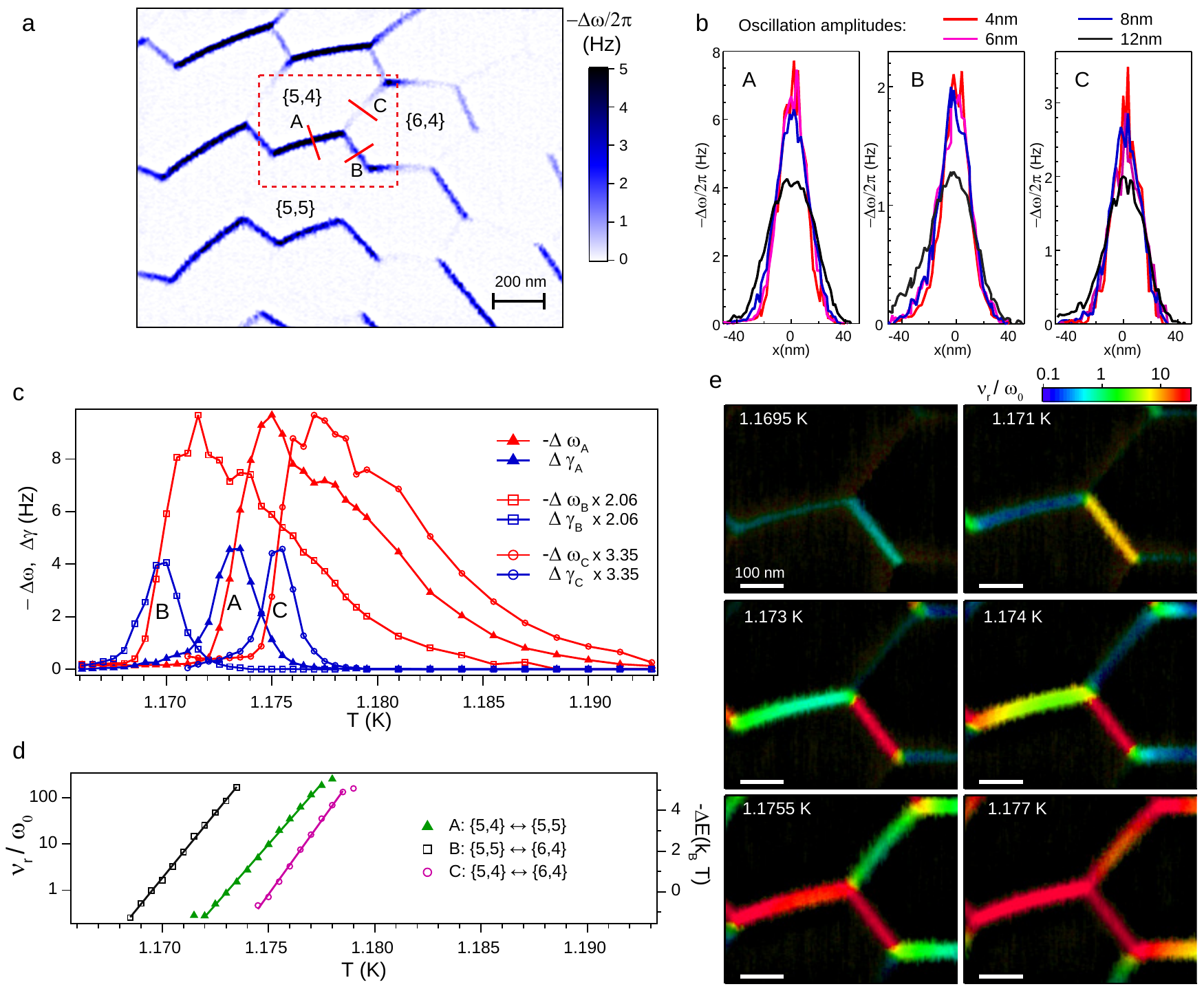}
\caption []{Temperature  dependent dynamics of the  transitions between  states $\{5,4\}$, $\{5,5\}$ and $\{6,4\}$.
(a) Positions of the line scans taken across three transitions are shown as solid red lines. 
(b)Resonance frequency shift measured for transitions A,B and C with several cantilever oscillations amplitude does not show  dependence  except at the highest amplitude of 12~nm.
(c) Temperature dependence  of the peak frequency (red) and  dissipation (blue) shifts of the cantilever.The relative strengths of the signal for  transitions A,B and C are scaled for better presentation. 
(d) Rate for different transitions, derived from $\Delta \omega$ and $\Delta \gamma$ shown in (C). The right axis shows the change of the phase slips energy barrier.
(e) Full stochastic resonance images  of the region shown in (a).}
\label{FigS5}
\end{figure*} 

Remarkably, the ratio $\Delta\omega/\Delta\gamma$, obtained from Eqs.~\eqref{eq:Frequency} and \eqref{eq:Damping}, has a simple form and gives a relaxation rate $\nu_r$ in units of cantilever frequency:
\begin{equation}\label{eq:phaseSlipRateS}
\frac{ \nu_r}{\omega_0}=\frac{ \Delta \omega}{\Delta \gamma}.
\end{equation}

Fig.~\ref{FigS5} shows data obtained for  vortex transitions between states $\{5,5\}$, $\{5,4\}$ and $\{6,4\}$.
The frequency shift data shown in Fig.~\ref{FigS5}b were measured by taking scans along  the lines marked as A, B, C in Fig.~\ref{FigS5}a.
All three transitions show  no dependence on the cantilever oscillation amplitude except at the  highest amplitude of 12~nm. Moreover, the relative frequency shifts are small ($\Delta \omega/\omega_0<0.002$).
Thus, the modulation is indeed sufficiently weak to justify the assumptions of our model of driven vortex transitions. Hence Eq.~\eqref{eq:phaseSlipRateS} can be used to extract the relaxation rates of the transitions.

The peak shifts in the  resonant frequency and the dissipation, which were extracted from the line scans, are shown in figure~\ref{FigS5}c as a function of temperature.
As can be seen from Fig.~\ref{FigS5}c, the signal on all three vortex transitions shows behavior consistent with stochastic resonance model given by Eqs.~\eqref{eq:Frequency} and \eqref{eq:Damping}: the dissipation $\Delta \gamma$ reaches a maximum (when $\nu_r\sim\omega_0$), while the resonant frequency shift $\Delta \omega$ grows rapidly; moreover, at this point $-\Delta \omega \simeq \Delta \gamma$.
The $\nu_r\sim\omega_0$ condition for transition C is observed around 1.175~K,  for transition A -- around 1.173~K, and for  transitions~B at 1.170~K.
The rates $\nu_r$, derived  for each transition using equation Eq.~\eqref{eq:phaseSlipRateS},  are plotted in figure~\ref{FigS5}d.
The right axis shows the relative change of the phase slip barrier heights in units of $k_B T$, which was calculated under the assumption that the attempt frequency does not change significantly within the temperature range of the measurements: $-\Delta E/k_BT \simeq \ln (\nu_r)$.
Figure~\ref{FigS5}e provides full images of $\nu_r$ for several temperatures in addition to those shown in Fig.~3 of the main text.
The images of the phase slip rate demonstrate a unique way  $\Phi_0$-MFM allows us to characterize  the dynamics of individual vortex transitions even in complex structures.

\end{document}